\documentclass[twocolumn,english,aps,superscriptaddress,prb,floatfix]{revtex4}
\usepackage[T1]{fontenc}
\usepackage[latin9]{inputenc}
\usepackage{xcolor}
\usepackage{amsmath}
\usepackage{graphicx}
\usepackage{amssymb}
\allowdisplaybreaks
\usepackage{epsf}
\usepackage{pinlabel}

\usepackage{bm}
\usepackage{bbm}
\usepackage[caption=false]{subfig}

\newcommand{\sio}{Sr$_2$IrO$_4$}

\newcommand{\bsio}{Sr$_3$Ir$_2$O$_7$}
\newcommand{\bslio}{(Sr$_{1-x}$La$_x$)$_3$Ir$_2$O$_7$}

\newcommand{\Ztwo}{\mathbb{Z}_2}

\newcommand{\HH}{\mathcal{H}}
\newcommand{\Ht}{\mathcal{H}_t}
\newcommand{\HJ}{\mathcal{H}_J}
\newcommand{\HJtwo}{\mathcal{H}_{J,2}}

\newcommand{\Heff}{\mathcal{H}_{\text{eff}}}

\newcommand{\Ac}{\mathcal{A}}

\newcommand{\corr}[1]{{\color{black}#1}}

\newcommand{\rme}{\mathrm{e}}

\newcommand{\rmi}{\mathrm{i}}
\newcommand{\rmd}{\mathrm{d}}
\renewcommand{\vec}[1]{\boldsymbol{#1}}
\newcommand{\mat}[1]{\bm{#1}}

\graphicspath{{../figs/}{./}{./figs/}{../v10/}}

\begin{document}

\title{
Carrier dynamics in doped bilayer iridates near magnetic quantum criticality
}

\author{Shouryya Ray}
\affiliation{Institut f\"ur Theoretische Physik, Technische Universit\"at Dresden,
01062 Dresden, Germany}
\author{Matthias Vojta}
\affiliation{Institut f\"ur Theoretische Physik, Technische Universit\"at Dresden,
01062 Dresden, Germany}
\affiliation{Center for Transport and Devices of Emergent Materials, Technische Universit\"at Dresden,
01062 Dresden, Germany}


\begin{abstract}
Motivated by experiments on the carrier-doped bilayer iridate {\bslio}, we study the dynamics of a single doped electron in a bilayer magnet in the presence of spin-orbit coupling, taking into account the spatially staggered rotation of IrO$_6$ octahedra. We employ an effective single-orbital bilayer $t$--$J$ model, concentrating on the quantum paramagnetic phase near the magnetic quantum critical point. We determine the carrier dispersion using a combination of self-consistent Born and bond-operator techniques. Extrapolating to finite small carrier density we find that, for experimentally relevant parameters, the combination of octahedral rotation and spin-orbit coupling induces a band folding which results in a Fermi surface of small double electron pockets, in striking agreement with experimental observations. We also determine the influence of spin-orbit coupling on the location of the quantum critical point in the undoped case, and discuss aspects of the global phase diagram of doped bilayer Mott insulators.
\end{abstract}

\date{4 September 2018}


\maketitle

\section{Introduction}

Iridium-oxide compounds form a fascinating class of materials, as their electronic properties are characterized by the simultaneous presence of strong electron--electron repulsion and strong spin-orbit coupling (SOC). A large body of work has been devoted to insulating iridates with Ir$^{4+}$ ions in a $5d^5$ electronic configuration which realize spin-orbit Mott insulators.\cite{kim08,moon08,kim09} These provide a fertile ground for novel forms of magnetism beyond that described by Heisenberg models.\cite{jin09,witczakkrempa14,machida10}
Moreover, in some of these materials, charge carriers have been successfully introduced by chemical doping, leading to an interplay of non-trivial magnetism and metallicity and associated insulator-to-metal transitions.\cite{li13,kim14,cao16} Such doped spin-orbit Mott insulators are not only interesting in their own right, but can also be expected to show parallels to cuprates which are well known for the emergence of high-temperature superconductivity.\cite{leermp} In fact, cuprate phenomenology holds a number of unsolved fundamental questions, and investigating other families of doped Mott insulators is therefore of broad interest.

The Ruddlesden--Popper series of iridates\cite{moon08} Sr$_{n+1}$Ir$_n$O$_{3n+1}$, with $n$ representing the number of square-lattice IrO$_2$ layers per unit cell, has been investigated in some detail over the past decade. The bilayer material {\bsio} is particularly fascinating for a number of reasons. First, it has been deduced from experimental data that it displays---in contrast to the bilayer cuprate YBa$_2$Cu$_3$O$_6$---a rather strong magnetic interlayer coupling which locates its magnetic ground state proximate to the magnetic quantum phase transition (QPT) known to exist for bilayer antiferromagnets.\cite{moretti15} Second, electron-doped {\bslio} shows an interesting evolution of magnetic, spectral, and transport properties.\cite{li13,torre14,hogan16,lu17}
In particular, a recent angle-resolved photoemission (ARPES) experiment\cite{torre14} has determined the low-energy electronic bands of weakly doped {\bslio}. The data indicates the presence of small double Fermi pockets with a momentum-space volume scaling with the doping level. We recall that Fermi pockets continue to be a source of debate in underdoped cuprates, as they have never been observed experimentally (in zero magnetic field) beyond doubt. Notably, {\bslio} displays two important differences compared to all cuprates, namely strong bilayer coupling and strong SOC.

This motivates us to study the dynamics of doped charge carriers in spin-orbit-coupled bilayer magnets in some detail---this is the purpose of our paper. To this end, we consider a suitable $t$--$J$ model and generalize earlier work on single-carrier dynamics \cite{vojtabecker98,holt12,holt13} to include effects of SOC. Since small doping pushes {\bslio} into the paramagnetic phase,\cite{li13,hogan16} we focus on the paramagnetic nearly critical regime of the undoped host magnet.
We derive an effective model for the carrier dynamics using generalized bond operators and determine the single-electron spectral function using a combination of bond-operator mean-field theory and self-consistent Born approximation. The combination of staggered rotation of IrO$_6$ octahedra and SOC leads to an enlarged unit cell and associated band folding independent of magnetic order, such that spectral features generically acquire ``shadows'' shifted by wavevector $(\pi,\pi,\pi)$. For realistic model parameters, we obtain a low-doping Fermi surface consisting of double pockets in agreement with experimental data. We also discuss broader aspects of doped bilayer Mott insulators.

The body of the paper is organized as follows:
Section~\ref{sec:model} introduces the microscopic modeling for {\bslio}. Section~\ref{sec:undoped} is devoted to the undoped system, discussing its bond-operator description and magnetic phase diagram. Section~\ref{sec:scba} describes the self-consistent Born approximation which we employ to determine the dynamics of doped electrons. Results for the latter are shown in Sec.~\ref{sec:res}, where in particular the momentum-dependent single-particle spectrum at low energies is displayed and compared to experimental data. A discussion and outlook will close the paper.


\section{Modeling}
\label{sec:model}

The active electrons in {\bslio} are those located in the Ir $5d$ orbitals. As has been discussed extensively, a combination of crystal-field and spin-orbit effects generates a Kramers doublet as the effective low-energy degree of freedom in the undoped system:\cite{kim08}
First, the five degenerate $d$ orbitals are split by the crystal field into a $t_{2g}$ triplet and a higher-lying $e_{g}$ doublet (each level with a spin multiplicity $2$). The latter, due to a significant separation $\sim 2\,\text{eV}$ from the $t_{2g}$ levels, are not part of a low-energy description. The $t_{2g}$ subspace is further split by strong SOC into a lower-lying $J_{\text{eff}} = 3/2$ quartet and a $J_\text{eff} = 1/2$ doublet, separated by an energy gap given by the SOC strength. Since the electron configuration of Ir$^{4+}$ is $5d^5$, the $J_\text{eff} = 3/2$ are completely filled, leaving a half-filled $J_\text{eff} = 1/2$ band. This band is finally rendered insulating by Hubbard repulsion, resulting in a $J_\text{eff} = 1/2$ spin-orbit Mott insulator.

The doping of additional electrons, as in {\bslio}, leads to the presence of $5d^6$ states, corresponding to a filled $t_{2g}$ multiplet. Below, we shall refer to $5d^6$ configurations of Ir ions as doublons.

\subsection{Pseudospin-$1/2$ $t$--$J$ model with spin-orbit anisotropy}

\begin{figure}
\includegraphics[width=.97\columnwidth]{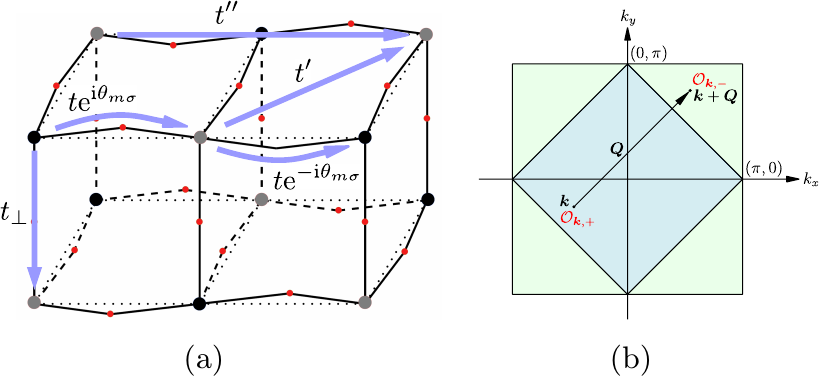}
\caption{
(a) Illustration of hopping paths in the bilayer structure of {\bslio}, with black and grey dots showing the two sublattices of Ir atoms and red dots showing oxygen locations.
(b) Brillouin zone (BZ) of the underlying square lattice, together with the reduced Brillouin zone $\text{BZ}^\prime$ corresponding to the doubled unit cell in Eq.~\eqref{ht}.
}
\label{fig:latt}
\end{figure}

Given the three relevant atomic states---the $5d^5$ pseudospin doublet and the $5d^6$ fully occupied singlet---a natural theoretical framework is an effective pseudospin-$1/2$ $t$--$J$ model on a bilayer square lattice,  ${\HH} = {\HH}_t + {\HH}_J$. Its hopping piece can be written in the form\cite{mei12}
\begin{align}
\label{ht}
{\HH}_t &= -t\!\!\sum_{\langle ii^\prime \rangle m\sigma} \rme^{\rmi \theta_{im\sigma}} \hat{c}^{\dagger}_{i m\sigma}\kern.05em\hat{c}_{i^\prime m\sigma} -t^\prime\!\!\! \sum_{\langle\!\langle ii^\prime \rangle\!\rangle m\sigma}\!\!\! \hat{c}^{\dagger}_{im\sigma}\kern.05em\hat{c}_{i^\prime m\sigma} \\
&\qquad -t^{\prime\prime}\!\!\!\! \sum_{\langle\!\langle\!\langle i i^\prime \rangle\!\rangle\!\rangle m\sigma}\!\!\!\! \hat{c}^{\dagger}_{im\sigma}\kern.05em\hat{c}_{i^\prime m\sigma} -t_\perp \sum_{i\sigma} \hat{c}^\dagger_{i1\sigma}\kern.05em\hat{c}_{i2\sigma} + \textnormal{H.c.}
\notag
\end{align}
The $\hat{c}_{im\sigma}$ are Gutzwiller-projected operators which annihilate an electron with pseudospin $\sigma=\uparrow,\downarrow$ at position $\vec{R}_i$ and layer $m=1,2$. In terms of canonical fermions $c_{im\sigma}$, these are given by $
\hat{c}^\dagger_{im\sigma} = c^\dagger_{im\sigma}\kern.05em n_{im\overline{\sigma}}$,
with $n = c^\dagger c$ (the overline on a binary index, such as $\sigma$ here, denotes its complement). The above construction, which excludes zero occupancy, is the appropriate one for the present case of electron doping.
The orbital content of the pseudospin states necessitates the consideration of the most general (with respect to pseudospin structure) time-reversal symmetric hopping bilinear.\cite{witczakkrempa14,fn:ba2iro4} Nearest-neighbor hopping is mediated via Ir-O-Ir bonds which are twisted in alternating directions due to the staggered rotation of the IrO$_6$ octahedra. The hopping bilinear then obtains an additional imaginary spin-dependent contribution, which can be subsumed---due to the still-maintained pseudospin-diagonal structure---into a phase factor of the hopping amplitude $t$.\cite{jin09,wangsenthil09,mei12}
It reads $\theta_{im\sigma} = \rme^{\rmi \vec{Q}\cdot \vec{R}_i} \eta_m \eta_\sigma \theta = \pm \theta$ whose sign involves the spin-orbit pseudospin $\eta_\sigma = \pm 1$ for $\sigma=\uparrow,\downarrow$, the sign of the layer pseudospin $\eta_m = (-1)^m$, and the sublattice sign $\rme^{\rmi \vec{Q}\cdot \vec{R}_i}$ with\cite{fn:units} $\vec{Q} = (\pi,\pi)$ due to the spatial pattern of the bond twisting, see Fig.~\ref{fig:latt}(a). The magnitude $\theta$ of the hopping phase parameterizes the anisotropy, and in particular vanishes for straight bonds. It need not, however, coincide (except in the simplest of cases) with the physical angle of octahedral rotation, but can attain corrections depending on the underlying $d$-orbital matrix elements.\cite{jin09}
Finally, $t'$ and $t''$ are the matrix elements for in-plane second- and third-neighbor hopping, while $t_\perp$ denotes the inter-layer hopping, see Fig.~\ref{fig:latt}(a).

As a consequence of the peculiar nearest-neighbor hopping matrix elements, non-Heisenberg interactions arise in the form of a pseudodipolar term with pseudodipolar tensor $\mat{\Gamma} = 2J\sin^2\theta\operatorname{diag}(0,0,1)$ and a Dzyaloshinskii--Moriya (DM) term with DM vector $\vec{D} = -J\sin(2\theta_{im})\kern.1em\vec{e}_z$.
The resulting magnetic piece of the Hamiltonian reads\cite{mei12}
\begin{align}
\HJ &= J\!\!\sum_{\langle ii^\prime \rangle m}\!\bigl(\cos(2\theta)\,\vec{S}_{im}\cdot \vec{S}_{i^\prime m} + 2\sin^2\theta\,S_{imz}\kern.05em S_{i^\prime mz} \nonumber \\[-.75em]       &\qquad\qquad\qquad -\sin(2\theta_{im})\,\vec{e}_z \cdot \vec{S}_{im} \times \vec{S}_{i^\prime m} \bigr) \\ &\qquad + J_\perp \sum_{i} \vec{S}_{i1}\cdot \vec{S}_{i2}\;, \nonumber
\end{align}
with $\theta_{im} = {\rme}^{\rmi \vec{Q}\cdot \vec{R}_i}\eta_m \theta$. The spin operators $\vec{S}_{im}$ in $\HJ$ are related to the canonical fermions in the usual manner via
\[
S_{im\alpha} = \tfrac{1}{2} c^\dagger_{im\sigma}\kern.05em\tau^\alpha_{\sigma\sigma^\prime}\kern.05em c_{im\sigma^\prime}\;.
\]
Note that we have neglected exchange interactions beyond nearest neighbors, i.e. the $J$-analogs of the $t^{\prime},t^{\prime\prime}$ terms from ${\HH}_t$, for two reasons: First, these terms are expected to be small (recall $J'/J\propto (t'/t)^2$ if derived from a Hubbard model), and second they will influence the carrier dynamics only indirectly (in contrast to $t'$ and $t''$).


\begin{figure}
\includegraphics[width=.75\columnwidth]{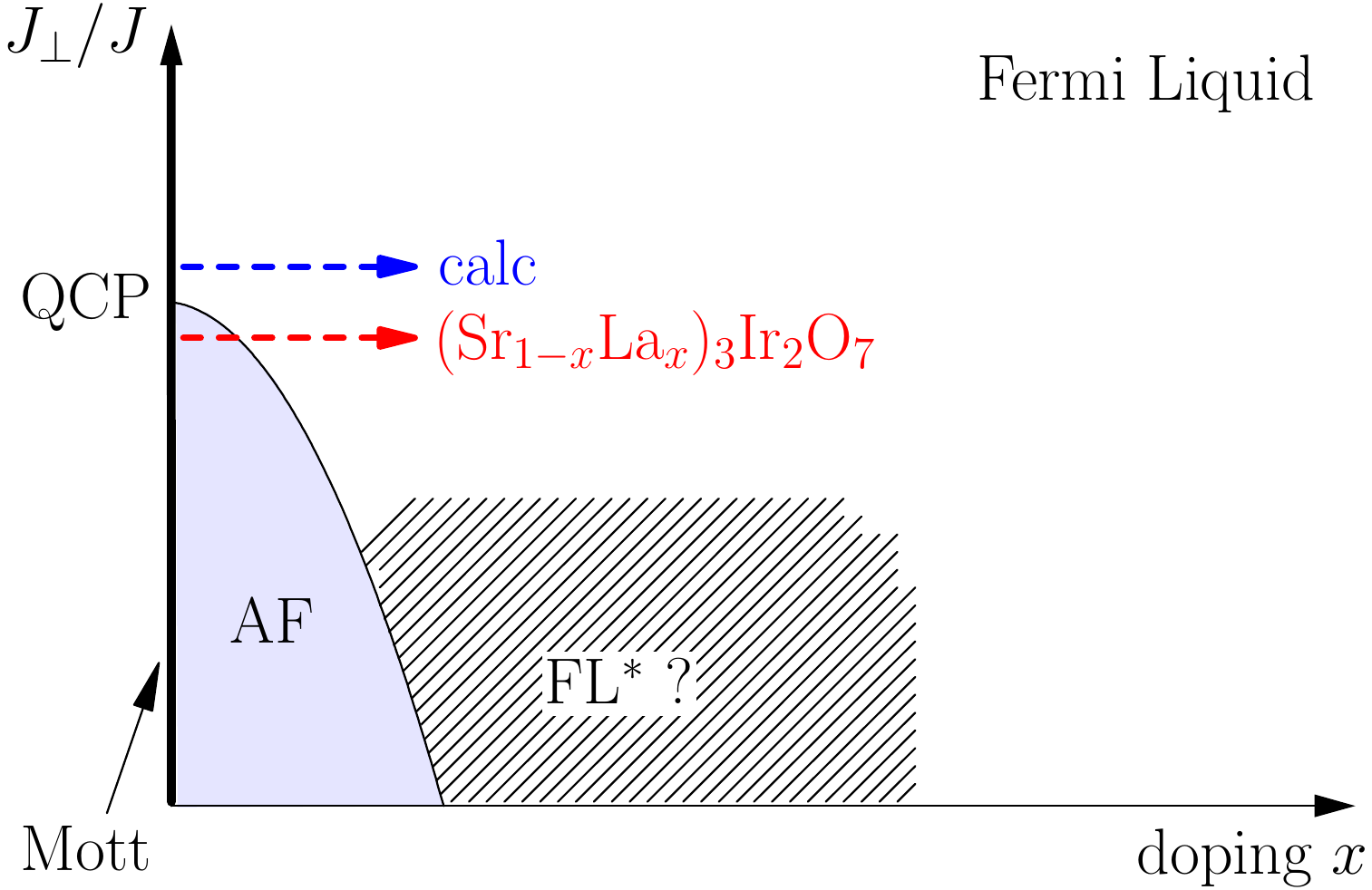}
\caption{
Schematic phase diagram of doped bilayer Mott insulators as function of 
carrier doping $x$ and magnetic interlayer coupling $J_\perp/J$. The 
undoped system can be tuned from an antiferromagnet (AF) to a dimer 
quantum paramagnet by increasing $J_\perp/J$. The dashed red line 
illustrates the path corresponding to \bslio; our calculation (dashed 
blue) considers carrier dynamics in the paramagnetic phase. The doped 
system both at large $J_\perp/J$ and at large $x$ is expected to be a 
Fermi liquid (FL). In contrast, the weakly doped paramagnetic single-layer 
Mott insulator likely is not. Instead it has been proposed to realize a 
fractionalized Fermi liquid (FL$^\ast$);\cite{flst1,moonss,ss16} this 
phase can be expected to continue to small finite $J_\perp$ (hatched). 
Additional symmetry-breaking instabilities (charge density wave, 
superconductivity, etc.) are not shown. }
\label{fig:schempd}
\end{figure}

\subsection{Choice of model parameters}
\label{sec:para}

While the above model is, to a certain degree, a generic model for doped bilayer square-lattice Mott insulators with spin-orbit coupling, our immediate goal is to describe the physics of {\bslio} at small $x$. We briefly outline the rationale behind the choice of parameters for our main calculations, noting that a complete set of microscopic parameters from first-principles calculations is, to the best of our knowledge, not available for \bsio{} which leads to some ambiguity.

First, we fix $J = 1$ as unit of energy. For \sio{}, we may use the values of hopping and interaction matrix elements specified in Ref.~\onlinecite{jin09} to extract a value of 2.5 for the ratio |t/J|. %
Since the bilayer iridates are known to be closer to the metal--insulator transition than their deeply insulating single-layer relatives, we assume a larger value of $t/U$. Since $t/J \propto\,U/t$ from the underlying Hubbard model, we take $t = -2$ as a rough estimate, where the minus sign is a consequence of the underlying band structure.\cite{plotnikova,fn:tsign} %
The longer-ranged hopping parameters $t^{\prime}$ and $t^{\prime\prime}$ have the same sign as $t$.\cite{wangsenthil09} Since the precise ratios depend sensitively on quantum chemistry, we treat them as virtually free parameters, and, for our main parameter set, tune them to reproduce the experimentally observed Fermi surface.
A reasonable combination was thus found to be $t^{\prime}/t = 0.35$ and $t^{\prime\prime}/t = 0.3$, which respects the ``natural'' hierarchy $|t| > |t^\prime| > |t^{\prime\prime}|$ expected based on the relative lengths of the hopping paths. For the anisotropy parameter, we put $\theta = 12^\circ$ to coincide with the physical rotation angle of the IrO$_6$ octahedra.

The final and important point is the choice of the interlayer coupling $J_\perp$. As discussed in detail below, increasing $J_\perp/J$ in the undoped case drives a QPT between an antiferromagnet and a quantum paramagnet, cf. Fig.~\ref{fig:schempd}. Generally, the bilayer coupling in {\bslio} is believed to be significantly larger than in bilayer cuprates, as signalled, e.g., by the large bilayer splitting seen in angle-resolved photoemission (ARPES).\cite{bil_split} Together with the data in Ref.~\onlinecite{moretti15}, this suggests that undoped {\bsio}---while antiferromagnetic---is presumably located close to the QPT. It becomes paramagnetic upon doping a small number ($x \approx 3$--$4\,\%$) of electrons, i.e., antiferromagnetism is destroyed by carrier motion.
In the calculational scheme we employ below, however, we neglect the feedback of carrier motion on magnetism for simplicity. Since we are interested in the physics of the paramagnetic phase, we choose $J_\perp/J$ such that it places the undoped system slightly into the paramagnetic phase, i.e., we assume that the feedback of carriers magnetism can be captured---to leading order---by an increase of $J_\perp/J$, see Fig.~\ref{fig:schempd}. In practice, we choose the interlayer coupling $J_\perp$ to yield a small triplon gap $\Delta = 0.2$, and assume the interlayer hopping amplitude to be positive,\cite{fn:tperpsign} with the absolute value in accordance with the Hubbard model ratio $(t_\perp/t)^2 = J_\perp/J \approx (J_\perp/J)_\text{c}$.

Due to the rough, sometimes \textit{ad hoc} nature of these estimates, we will also take the liberty of exploring neighboring regions of parameter space by illustrating general trends regarding the evolution of the fermiology upon changing individual parameters.


\section{Undoped system: Bond operators and magnetic QPT}
\label{sec:undoped}

We first turn our attention to the half-filled case, where we have the Hamiltonian $\HJ$ involving spin degrees of freedom only. It is known\cite{chub95,kotovsushkov98,matsushita99} that such bilayer antiferromagnetic Heisenberg models possess a quantum critical point (QCP) tuned by $J_\perp/J$, separating an interlayer dimerized paramagnetic state ($J_\perp/J \gg 1$) from an intralayer N{\'e}el antiferromagnet ($J_\perp/J \ll 1$) with ordering wavevector ${\vec Q} = (\pi,\pi)$. We note that for small $J_\perp$, the combination of pseudodipolar and DM interactions leaves the collinear N\'eel state intact, as opposed to the case of only DM interaction which would induce spiral order.

For an efficient description we employ a variant of bond-operator theory, originally due to Sachdev and Bhatt,\cite{sachdevbhatt90} generalized to include SOC effects. Compared to plain spin-wave theory, the bond-operator approach has the advantage of capturing longitudinal fluctuations (i.e. the Higgs mode) which is important near the QCP and which has been argued to be present in \bsio.\cite{moretti15}

\subsection{Bond-operator representation}

At each dimer site $i$, the total spin eigenbasis $\{|t_{i\mu} \rangle ; \mu = 0,\ldots,3\}$ of the Hilbert space of two dimer spins comprises a singlet $|t_{i0} \rangle$ and a triplet $|t_{i\alpha}\rangle$:
\begin{align*}
|t_{i0}\rangle &= \frac{1}{\sqrt{2}}\left({c}_{i1\uparrow}^\dagger\kern.05em {c}_{i2\downarrow}^\dagger - {c}_{i1\downarrow}^\dagger\kern.05em {c}_{i2\uparrow}^\dagger\right)\bigl|0\bigr\rangle, \\
|t_{i1}\rangle &= \frac{-1}{\sqrt{2}}\left({c}_{i1\uparrow}^\dagger\kern.05em {c}_{i2\uparrow}^\dagger - {c}_{i1\downarrow}^\dagger\kern.05em {c}_{i2\downarrow}^\dagger\right)\bigl|0\bigr\rangle, \\
|t_{i2}\rangle &= \frac{\rmi}{\sqrt{2}}\left({c}_{i1\uparrow}^\dagger\kern.05em {c}_{i2\uparrow}^\dagger + {c}_{i1\downarrow}^\dagger\kern.05em {c}_{i2\downarrow}^\dagger\right)\bigl|0\bigr\rangle, \\
|t_{i3}\rangle &= \frac{1}{\sqrt{2}}\left({c}_{i1\uparrow}^\dagger \kern.05em {c}_{i2\downarrow}^\dagger + {c}_{i1\downarrow}^\dagger \kern.05em {c}_{i2\uparrow}^\dagger\right)\bigl|0\bigr\rangle,
\end{align*}
where $|0\rangle$ is the electron vacuum, i.e. $c_{im\sigma}|0\rangle = 0$. To proceed, one introduces bosonic bond operators that create these states out of a fictitous vacuum, $|t_{i\mu}\rangle = t_{i\mu}^\dagger|\text{vac}\rangle$. Since the physical states are either singlets or triplets, the constraint
\begin{align}
\sum_{\mu=0}^3 t_{i\mu}^\dagger \kern.05em t_{i\mu} = 1
\label{eq:triplconstr}
\end{align}
has to be fulfilled. The spin operators can then be represented as (site index $i$ suppressed for brevity):
\begin{align}
S_{m\alpha} = \frac{1}{2} \left[-\eta_m \bigl(t_0^\dagger\kern.05em t_\alpha + t_\alpha^\dagger\kern.05em t_0\bigr) - \sum_{\beta\gamma}\rmi\kern.05em\epsilon_{\alpha \beta \gamma} t_\beta^\dagger\kern.05em t_\gamma\right]
\label{eq:bondrep}
\end{align}
for $\alpha=1,2,3\equiv x,y,z$. Insertion into ${\HH}_J$ yields a Hamiltonian with upto four-body interactions, turning ${\HH}_J$ into a problem of interacting hard-core bosons.


\subsection{Self-consistent mean-field theory}

For a simple yet quantitative treatment of ${\HH}_J$ in the paramagnetic phase, we resort to bond-operator mean-field (BOMF) theory.\cite{sachdevbhatt90} For the $SU(2)$-symmetric bilayer Heisenberg model this has been presented in Ref.~\onlinecite{matsushita99}, and we generalize their BOMF solution to include effects of SOC and bond twisting. We note that BOMF is quantitatively more accurate than the simplest harmonic approximation to bond-operator theory, the latter ignoring the constraint altogether, see Sec.~\ref{sec:gapqcp}.

As the paramagnetic phase is adiabatically connected to a product state of dimer singlets, it is convenient to condense the singlets, $t_{i0} \to \langle t_{i0}\rangle \equiv s \in \mathbb{R}$. The constraint \eqref{eq:triplconstr} is enforced on average using a Lagrange multiplier $\lambda$, i.e.,
\[
\HJ \to \HJ - \lambda\sum_{i}\left(s^2 + \sum_\alpha t_{i\alpha}^\dagger \kern.05em t_{i\alpha} - 1\right).
\]
Neglecting all triplon--triplon interactions, the resulting quadratic Hamiltonian in momentum space is given by\cite{fn:morettisalaHO}
\begin{align*}
\HJtwo &= \tfrac{1}{2}\sum_{\vec{q} \alpha} \left(t_{\vec{q}\alpha}^\dagger\;t_{-\vec{q},\alpha}\right)
\!
\left(
\begin{array}{ll}
A_{\vec{q}\alpha} & B_{\vec{q}\alpha} \\
B_{\vec{q}\alpha} & A_{\vec{q}\alpha}
\end{array}
\right)
\!
\left(\!
\begin{array}{l}
t_{\vec{q}\alpha} \\[.1em]
t_{-\vec{q},\alpha}^\dagger
\end{array}
\!
\right) \\
& \qquad - \tfrac{1}{2}\sum_{\vec{q}\alpha} A_{\vec{q}\alpha} + N\Bigl[\lambda - \left(\tfrac{3}{4}J_\perp + \lambda\right)\!s^2\Bigr]
\end{align*}
where
\begin{align}
\begin{split}
B_{\vec{q}\alpha} &= 2 J_\alpha s^2 \gamma_{\vec{q}}\;, \\
A_{\vec{q}\alpha} &= \tfrac{1}{4}J_\perp - \lambda + B_{\vec{q}\alpha}\;, \\
2\gamma_{\vec{q}} &= \cos q_x + \cos q_y\;,
\end{split}
\label{eq:HJtwohelpers}
\end{align}
with the effective polarization-dependent coupling $J_\alpha = J\bigl(\cos(2\theta) + 2\sin^2\theta\kern.05em\delta_{\alpha,z}\bigr)$. At this level of approximation, the effect of SOC is to reduce the symmetry of the triplon modes from $\text{O}(3)$ to $\Ztwo \times \text{O}(2)$. We note that contributions from the DM interaction cancel in $\HJtwo$ because of its antisymmetry. Consequently, the momentum summations in $\HJtwo$ are over the full Brillouin zone.
The Hamiltonian $\HJtwo$ is readily diagonalized using a Bogoliubov transformation to yield
\begin{align}
\begin{split}
\HJtwo &= \sum_{\vec{q}\alpha} \omega_{\vec{q}\alpha} \kern.05em \Bigr(\beta_{\vec{q}\alpha}^\dagger\kern.05em\beta_{\vec{q}\alpha} + \tfrac{1}{2}\Bigl) -\tfrac{1}{2}\sum_{\vec{q}\alpha}A_{\vec{q}\alpha} \\ &\qquad + N\Bigl[\lambda - \left(\tfrac{3}{4}J_\perp + \lambda\right)\!s^2\Bigr]\;,
\end{split}
\label{eq:HJtwo}
\end{align}
where
\begin{align}
\begin{split}
t_{\vec{q}\alpha} &= u_{\vec{q}\alpha} \kern.05em \beta_{\vec{q}\alpha} + v_{\vec{q}\alpha} \kern.05em \beta_{-\vec{q},\alpha}^\dagger, \\
u_{\vec{q}\alpha} &= \sqrt{\frac{A_{\vec{q}\alpha}}{2\omega_{\vec{q}\alpha}} + \frac{1}{2}}, \\
v_{\vec{q}\alpha} &= -\operatorname{sgn}\gamma_{\vec{q}}\sqrt{\frac{A_{\vec{q}\alpha}}{2\omega_{\vec{q}\alpha}} - \frac{1}{2}}, \\
\omega_{\vec{q}\alpha} &= \sqrt{A_{\vec{q}\alpha}^2 - B_{\vec{q}\alpha}^2}.
\end{split}
\label{eq:Bogovars}
\end{align}
The mean-field parameters can be obtained by minimizing the ground-state energy density, which is given by
\begin{align}
\frac{E_0}{N} = -\frac{1}{2N}\sum_{\vec{q}\alpha}^{}\left( A_{\vec{q}\alpha} - \omega_{\vec{q}\alpha} \right) + \lambda - \left(\tfrac{3}{4}J_\perp + \lambda\right)\!s^2\,.
\end{align}
The resulting self-consistency equations read:
\begin{align}
\begin{split}
s^2 &= \frac{5}{2} - \frac{1}{N}\sum_{\vec{q}\alpha}^{}\frac{1+(J_\alpha/J)d\gamma_{\vec{q}}}{2\sqrt{1+2(J_\alpha/J)d\gamma_{\vec{q}}}}\\
\lambda &= -\frac{3}{4}J_\perp + \frac{1}{N}\sum_{\vec{q}\alpha}\frac{J_\alpha\gamma_{\vec{q}}}{\sqrt{1+2(J_\alpha/J)d\gamma_{\vec{q}}}}
\end{split}
\end{align}
where we have introduced above, following Matsushita \textit{et al.},\cite{matsushita99} the combination
\begin{align}
d = 2Js^2\kern.05em(J_\perp/4 - \lambda)^{-1}\;.
\label{ddef}
\end{align}
This is particularly convenient, because it reduces the pair of self-consistency equations to one for $d$, viz.:
\begin{align}
d &= \frac{J}{J_\perp}\left(5 - \frac{1}{N}\sum_{\vec{q}\alpha}^{}\frac{1}{\sqrt{1+2(J_\alpha/J)d\gamma_{\vec{q}}}}\right).
\end{align}
The equation for $d$ is readily solved numerically and can then be used to determine $s,\lambda$ using the relations above. Thence, all other quantities of interest can be calculated.
In the limit $J_\perp/J \to \infty$, one finds $s \to 1$ and $\lambda \to -\tfrac{3}{4}J_\perp$, in which case the self-consistent mean-field treatment coincides with the harmonic approximation.\cite{fn:morettisalaHO}

\subsection{Triplon gap and location of QCP}
\label{sec:gapqcp}

\begin{figure}
\includegraphics[height=.23\textheight]{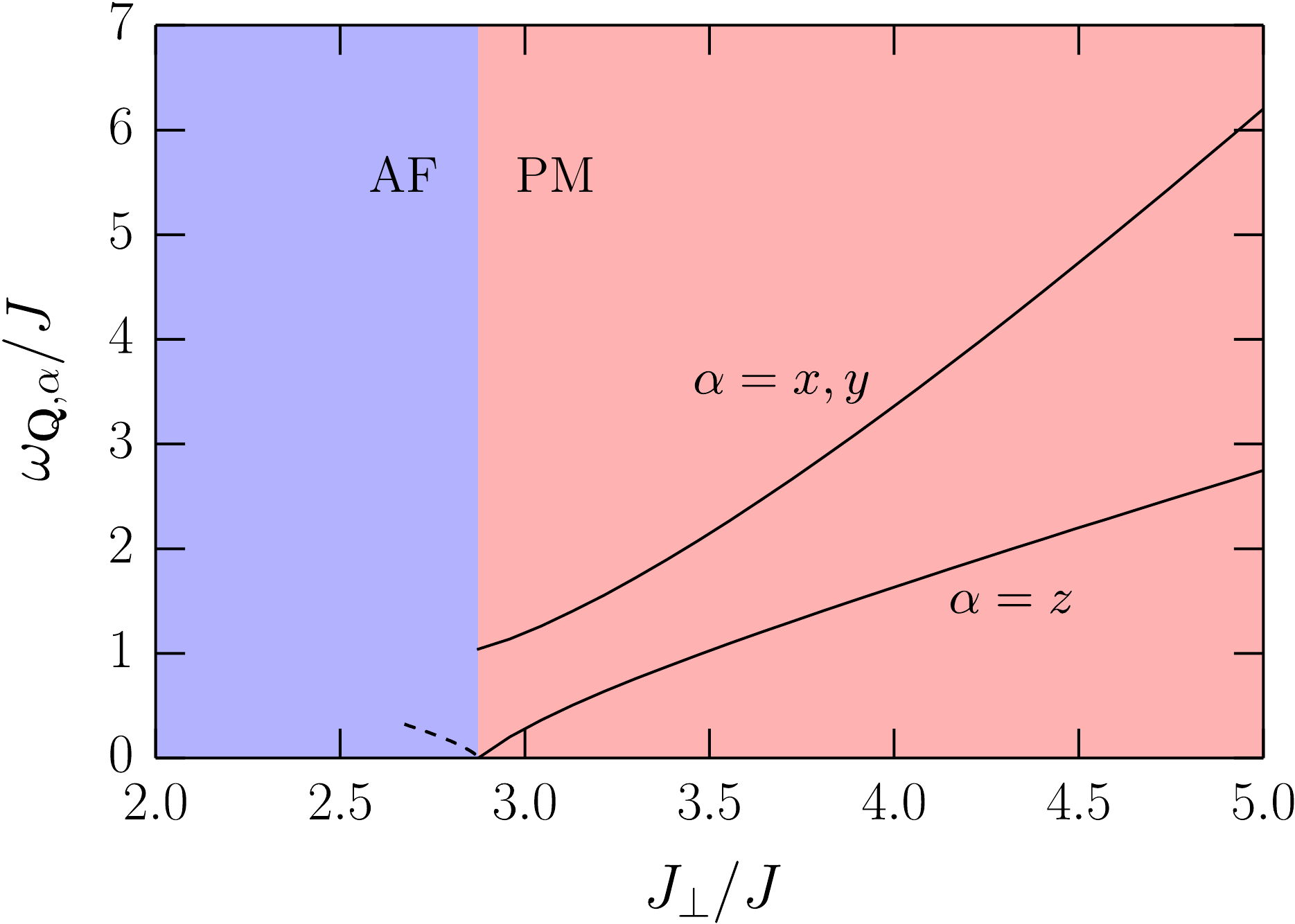}
\caption{
Minimum energy $\omega_{\vec{Q},\alpha}$ of the triplon dispersion for the two degenerate $x,y$ modes and the separate $z$ mode for $\theta = 12^\circ$. The latter goes soft first and hence controls the QPT.
}
\label{fig:dispmin}
\end{figure}

\begin{figure}[]
\includegraphics[height=.24\textheight]{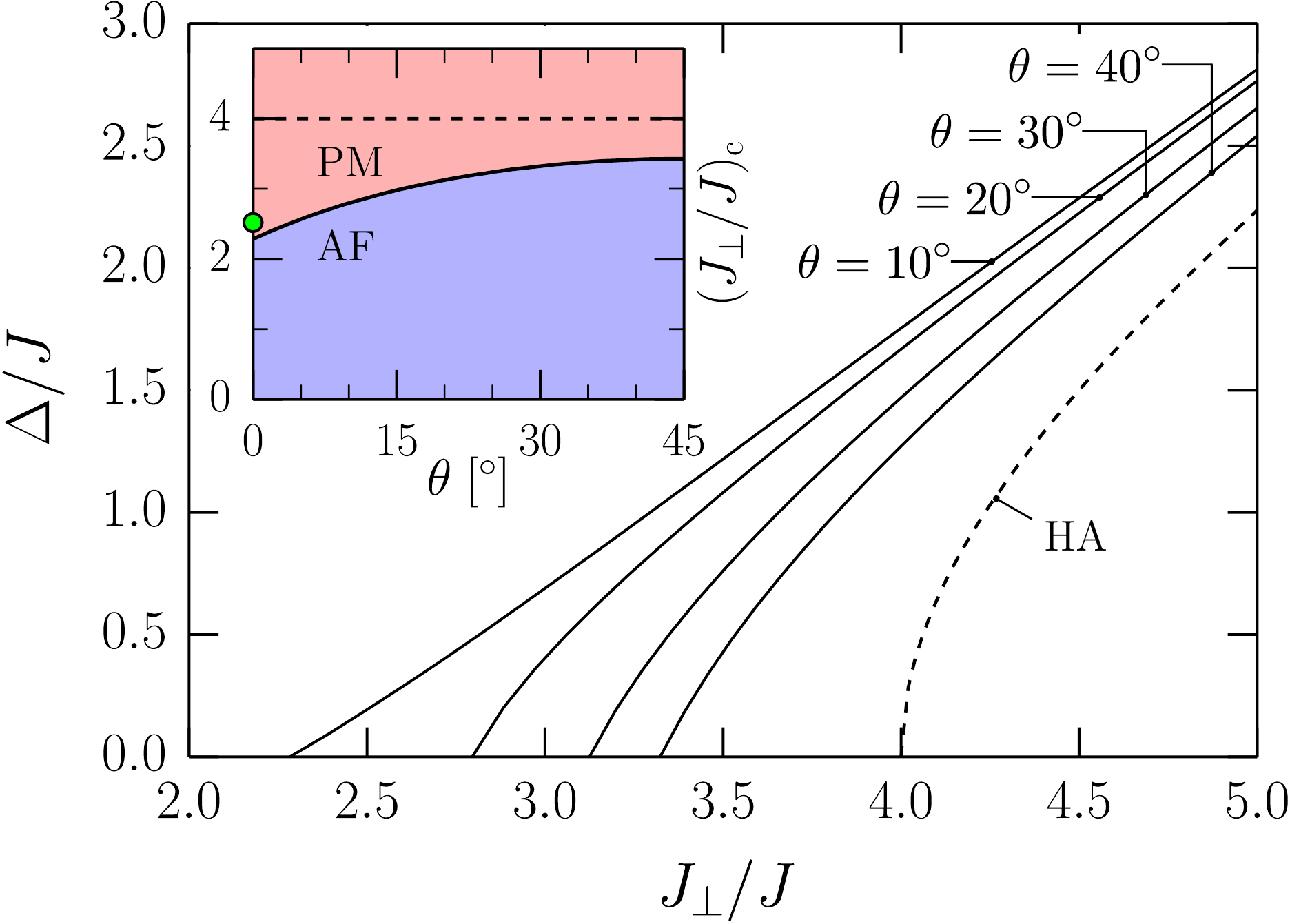}
\caption{
Evolution of the triplon gap $\Delta$ as a function of the tuning parameter $J_\perp$ from BOMF theory (solid) for different values of the SOC parameter $\theta = 10^\circ, 20^\circ, 30^\circ$ and $40^\circ$, along with the ($\theta$-independent) result obtained in harmonic approximation (dashed). The inset shows the location of the QCP as a function of $\theta$ from BOMF (solid) and harmonic (dashed) approximations. The green dot indicates the QMC result for $\theta = 0$ from Ref.~\onlinecite{sandvikscalapino94}.
}
\label{fig:Delta}
\end{figure}

Due to the combination of SOC and octahedral rotation, the three-fold degeneracy of the triplon modes is lifted. The triplon dispersions have the same shape as in the $\theta=0$ case (which would correspond to vanishing SOC or vanishing octahedral rotation), but with effective $J_\alpha$ for the respective polarizations rather than a common $J$. Thus, there is a pair of degenerate modes ($\alpha = x,y$) and a separate mode for $\alpha = z$. For all $\alpha$ the dispersion minimum lies at the putative ordering wavevector $\vec{Q} = (\pi,\pi)$ and is given by
\begin{align}
\omega_{\vec{Q},\alpha} = \left(\tfrac{1}{4}J_\perp - \lambda\right)\!\sqrt{1 - 2d\kern.05em (J_\alpha/J)}
\end{align}
with $d$ from Eq.~\eqref{ddef}.
Since $J_z = J$ is the largest effective coupling, the $z$ mode has the smallest gap and thus controls the QPT, as in the harmonic approximation.\cite{moretti15} The situation is illustrated for $\theta = 12^\circ$ in Fig.\ \ref{fig:dispmin}.
The minimum triplon gap $\Delta$ is given by
\begin{align}
\Delta = \omega_{\vec{Q},z} = \left(\tfrac{1}{4}J_\perp - \lambda\right)\!\sqrt{1 - 2d}\;,
\label{eq:QCPloc}
\end{align}
which we plot in Fig.\ \ref{fig:Delta} as a function of the tuning parameter $J_\perp$ for various values of $\theta$.

The condition $\Delta=0$ \eqref{eq:QCPloc} defines the QCP, and we plot the corresponding coupling $(J_\perp/J)_{\text{c}}$ as function of $\theta$ in the inset of Fig.~\ref{fig:Delta}.
Numerically accurate results, such as those obtained using Quantum Monte Carlo techniques,\cite{sandvikscalapino94} are available only for the SU(2)-invariant case and place the QCP at $(J_\perp/J)_{\text{c}} \approx 2.51$, so that the corresponding BOMF result $(J_\perp/J)_{\text{c}} \approx 2.28$ is remarkably close, especially when taking the simplicity of the method into account. In particular, it is a considerable improvement over the harmonic approximation, which neglects the constraint \eqref{eq:triplconstr} and yields a $\theta$-independent $(J_\perp/J)_{\text{c}} = 4$. We expect BOMF to yield quantitatively reasonable results for $\theta \neq 0$ as well.


\section{Carrier doping and self-consistent Born approximation}
\label{sec:scba}

To describe electron doping, we extend the dimer Hilbert space comprising the four two-electron states $|t_{i\mu}\rangle$ by an additional three-electron state $|d_{im\sigma}\rangle$. The quantum numbers $m\sigma$ refer to the electron needed to make the dimer fully occupied. We call the state a \textit{doublon}, because the site at position $\vec{R}_i$ and layer $\overline{m}$ is doubly occupied. We introduce a pseudofermion operator that creates the new doublon state out of the fictitous vacuum, i.e.
\begin{align*}
d_{im\sigma}^{\kern.05em\dagger}\kern.05em |\text{vac}\rangle = |d_{im\sigma}\rangle\;.
\end{align*}
Since the physical Hilbert space is restricted to the states $\{|t_{i\mu}\rangle,|d_{im\sigma}\rangle\}$, the extended constraint reads
\begin{align}
\sum_\mu t_{i\mu}^\dagger\kern.05em t_{i\mu} + \sum_{m\sigma} d_{im\sigma}^{\kern.05em\dagger}\kern.05em d_{im\sigma} = 1.
\end{align}
In direct analogy with the holon pseudofermion formalism,\cite{jureckabrenig01} we can now write down the action of the $\hat{c}$-operators within the aforementioned physical Hilbert space in terms of doublons and bond bosons as follows (spectator index $i$ suppressed for brevity):
\begin{align}
\hat{c}_{m\sigma}^\dagger = \frac{\eta_m}{\sqrt{2}}\left[d^{\kern.05em\dagger}_{\overline{m}\kern.05em\overline{\sigma}} \left(-\eta_m\eta_\sigma t_0 + t_3\right) + d^{\kern.05em\dagger}_{\overline{m}\sigma}\left(\eta_\sigma t_1 + \rmi t_2\right)\right].
\label{eq:pseudofrep}
\end{align}
Since the Hamiltonian $\HH$ contains explicit appearances of the sign of the sublattice (and therefore effectively doubles the unit cell), we promote every operator to a corresponding bipartite version. Formally, for an arbitrary operator $\mathcal{O}$, we define two new operators $\mathcal{O}_A$ and $\mathcal{O}_B$ defined only on the $A$ sublattice, and set
\begin{align}
\mathcal{O}_i =
\begin{cases}
\mathcal{O}_{iA} & i \in A \\
\mathcal{O}_{i-\vec{\delta},B} & i \in B
\end{cases}
\end{align}
where $\vec{\delta} = (1,0)$. The respective Fourier transforms are performed for momenta $\vec{k}$ in the reduced Brillouin zone $\text{BZ}^\prime$, Fig.~\ref{fig:latt}(b), as
\begin{align}
\begin{split}
\mathcal{O}_{\vec{k}A} &= \sqrt{\frac{2}{N}} \sum_{i\in A}\rme^{\rmi \vec{k}\cdot \vec{R}_i} \mathcal{O}_{iA}, \\
\mathcal{O}_{\vec{k}B} &= \sqrt{\frac{2}{N}} \sum_{i\in A}\rme^{\rmi \vec{k}\cdot (\vec{R}_i + \vec{\delta})} \mathcal{O}_{iB}.
\end{split}
\end{align}
Furthermore, we introduce $\mathcal{O}_{\kappa} \equiv (1/\sqrt{2})\left(\mathcal{O}_A + \kappa \mathcal{O}_B\right)$ where $\kappa = \pm 1$; for $\vec{k} \in \text{BZ}^\prime$ we have $\mathcal{O}_{\vec{k} + \vec{Q},\kappa} = \mathcal{O}_{\vec{k},-\kappa}$.

To derive the effective Hamiltonian governing the dynamics of doublons, we have to consider contributions from $\Ht$ and $\HJ$. The former simply entails the direct insertion of \eqref{eq:pseudofrep} into $\Ht$. For the latter, one has to extend the bond-boson representation \eqref{eq:bondrep} by including terms of the form $\langle d_{m_1\sigma_1} | S_{m\alpha} | d_{m_2\sigma_2} \rangle \kern.1em d_{m_1\sigma_1}^{\kern.05em\dagger} d_{m_2\sigma_2} = -\tfrac{1}{2} \tau^\alpha_{\sigma\sigma^\prime} d_{m\sigma^\prime}^{\kern.05em\dagger} d_{m\sigma}$ before inserting into $\HJ$.
This results in free doublon hopping and many-body interaction terms. Of the latter, we only keep the minimal interaction vertex consisting of one triplon and two doublon operators. Writing everything down in terms of bipartite operators $d_{i{\kappa}m\sigma},t_{i{\kappa}\alpha}$ and Fourier transforming, one obtains the final effective Hamiltonian
\begin{align}
\Heff &= \HH_{\psi\psi} + \HH_{\beta\beta} + \HH_{\beta\psi\psi}, \label{eq:efftheor0}\\
\HH_{\beta\beta} &= \sum_{\vec{q}{\kappa}\alpha}\omega_{\vec{q}{\kappa}\alpha}\kern.05em\beta_{\vec{q}{\kappa}\alpha}^\dagger\kern.05em\beta_{\vec{q}{\kappa}\alpha}, \\
\HH_{\psi\psi} &= \sum_{\vec{k}\sigma} \psi_{\vec{k}\sigma}^\dagger \kern.05em h_{\vec{k}\sigma} \kern.05em \psi_{\vec{k}\sigma}, \label{eq:efftheor-0}\\
\HH_{\beta\psi\psi} &= \sqrt{\frac{2}{N}}\sum_{\vec{q}\kappa\alpha} \sum_{\vec{k}\sigma\sigma'} \beta_{\vec{q}{\kappa}\alpha}\kern.05em \psi^\dagger_{\vec{k}+\vec{q},\sigma}\kern.05em g_{\vec{k}\vec{q}\kern.05em {\kappa} \alpha\sigma\sigma^\prime} \psi_{\vec{k}\sigma^\prime} + \textnormal{H.c.},
\label{eq:efftheor-1}
\end{align}
with all momentum sums restricted to the reduced Brillouin zone $\text{BZ}^\prime$. We have dropped the constant term and suitably promoted the triplon operators and other related quantities from $\HJtwo$ to their bipartite version.\cite{fn:triplonbip} Furthermore, we have introduced the $4$-spinor
\[
\psi_{\vec{k}\sigma}^\dagger = \left(d_{\vec{k},+,1\sigma}^{\kern.05em\dagger},d_{\vec{k},-,1\sigma}^{\kern.05em\dagger},d_{\vec{k},+,2\sigma}^{\kern.05em\dagger},d_{\vec{k},-,2\sigma}^{\kern.05em\dagger}\right)
\]
with the shorthand $\kappa=+,-$.
The Hamiltonian matrix $h_{\vec{k}\sigma}$ and the interaction vertices $g_{\vec{k}\vec{q}\kern.05em\alpha {\kappa}\sigma\sigma^\prime}$ are $4 \times 4$ matrices acting in spinor space, the (somewhat lengthy) explicit expressions for which are relegated to Appendix \ref{app:expl}. Importantly, the coupling between the $\kappa=+,-$ sectors reflects the reduced translational symmetry arising from the combination of octahedral rotation and SOC; it couples momenta $\vec{k}$ and $\vec{k}+\vec{Q}$ at the single-particle level and leads to an associated backfolding of bands.

With the effective theory \eqref{eq:efftheor0}--\eqref{eq:efftheor-1} at hand, we can now turn our attention towards the main goal of this study: charge carrier dynamics. To this end, we compute the (retarded) doublon Green's function, which is defined in the usual manner as
\begin{align}
G_\sigma(\vec{k},E) &= -\rmi \int\! \rmd t\,\rme^{\rmi E t}\kern.05em\Theta(t) \langle \Omega | \{ \psi_{\vec{k}\sigma}(0) , \psi_{\vec{k}\sigma}^\dagger(t) \} | \Omega \rangle \nonumber\\
&\equiv \langle\!\langle \psi_{\vec{k}\sigma} ; \psi^{\dagger}_{\vec{k}\sigma} \rangle\!\rangle_E
\label{gdoublon}
\end{align}
where $|\Omega\rangle$ is the half-filled ground state (\textit{per constructionem} a vacuum for both $\beta$-excitations and doublons). At tree level, the matrix propagator can be directly read off from $\HH_{\psi\psi}$:
\begin{align}
G_\sigma^{\kern.05em(0)}(\vec{k},E) = \frac{1}{ E + \rmi 0^+ - h_{\vec{k}\sigma} }
\end{align}
which is $\sigma$-diagonal, but depends on $\sigma$.
When evaluating the self-energy $\Sigma_\sigma(\vec{k},E)$, bare perturbation theory in the interaction strength is inadequate, because the interaction vertices are typically larger than unity (since $t/J \gg 1$). In such scenarios, the so-called \textit{self-consistent Born approximation} (SCBA) has established itself as a method of choice.\cite{schmittrink88,kaneleeread89,eder98,holt12,holt13} The key idea is to resum a certain subset of diagrams, namely those consisting of non-crossing triplon loops (hence also called \textit{non-crossing approximation}, NCA). This has to be done implicitly, and leads to the self-energy\cite{altlandsimons}
\begin{align}
\Sigma_\sigma(\vec{k},E) &= \frac{2}{N}\!\sum_{\vec{q}{\kappa}\alpha\sigma^\prime}\! g_{\vec{k}-\vec{q},\vec{q}\kern.05em{\kappa}\alpha\sigma\sigma^\prime} G_{\sigma^\prime}\bigl(\vec{k}-\vec{q},E - \omega_{\vec{q}{\kappa}\alpha}\bigr)
\times{}\nonumber \\[-0.25em] & \qquad\qquad\quad {} \times
g_{\vec{k}-\vec{q},\vec{q}{\kappa}\alpha\sigma^\prime\sigma}^{\kern.05em\dagger}\;,\label{eq:SigmaSCBA}
\end{align}
which is $\sigma$-diagonal due to a residual spin symmetry of the theory.\cite{fn:structureG} Thus, one needs to solve the fixed-point Dyson equation
\begin{align}
G_\sigma(\vec{k},E) = \frac{1}{G_\sigma^{\kern.05em(0)}(\vec{k},E)^{-1}_{\vphantom{\sigma}} - \Sigma_\sigma[G_\sigma](\vec{k},E)}\;,
\label{eq:Dyson}
\end{align}
to which end we proceed iteratively. For our main calculations, we solve the system \eqref{eq:SigmaSCBA}--\eqref{eq:Dyson} on a grid of $N = 32\sqrt{2} \times 32\sqrt{2}$ $k$-points, frequency resolution $\Delta E = 0.01$ and artificial broadening $\eta = 0.01$ (all energies in units of $J$).


\section{Doublon dynamics}
\label{sec:res}

With the formalism at hand, we now discuss numerical results and relate them to physical observables.

\subsection{Doublon propagators and photoemission}

Given that $\psi$ is an auxiliary pseudofermion, we start by relating its propagator to that of actual electrons. The standard observable (e.g. for photoemission) is the single-particle spectral function
\begin{align}
{\Ac}_{mm^\prime\sigma}(\vec{k},E) = -(1/\pi)\operatorname{Im}\langle\!\langle c_{\vec{k}m\sigma} ; c^{\dagger}_{\vec{k}m^\prime\sigma} \rangle\!\rangle_E\;,
\end{align}
where the 2-momentum $\vec{k}$ now resides in the full BZ, and we have assumed a spin-diagonal structure. Close to half-filling, the low-energy contributions to $\Ac$ come from the dynamics of doubly occupied sites, such that we may replace the $c_{\vec{k}m\sigma}$ by the projected operators $\hat{c}_{\vec{k}m\sigma}$ (this corresponds to ignoring excitations in the Hubbard bands). Finally, from the representation \eqref{eq:pseudofrep} in terms of doublon pseudofermions, we see that
\begin{align}
\langle\!\langle c_{\vec{k}m\sigma} ; c^{\dagger}_{\vec{k}m^\prime\sigma} \rangle\!\rangle_E = \frac{s^2}{2}\langle\!\langle {d}_{\vec{k}\overline{m}\kern.1em\overline{\sigma}} ; {d}^{\kern.1em\dagger}_{\overline{m^\prime}\kern.1em\overline{\sigma}}(K) \rangle\!\rangle_E + \cdots\;,\label{eq:doublon_to_obs}
\end{align}
where $\cdots$ contains (with suitable prefactors) composite Green's functions of the form
\[\langle\!\langle \bigl(d_{\vec{k}m_1\sigma_1} t_{\vec{q}\alpha_1}\bigr) ; \bigl(d^{\kern.1em\dagger}_{\vec{k}^\prime m_2 \sigma_2} t_{\vec{q}^\prime\alpha_2}\bigr) \rangle\!\rangle_E\;.\]
Such corrections contribute a continuum to the incoherent part of the spectral function only.
More importantly, they are suppressed by factors $v_{\vec{q}}$, as they involve the annihilation of a triplon.\cite{kotovsushkov98}
Hence we neglect them in the following. The most important contribution is thus found to be
\begin{align}
{\Ac}_{mm^\prime\sigma}(\vec{k},E) = -\frac{1}{\pi}\operatorname{Im} \left\{\begin{array}{ll}
\frac{s^2}{2}G_{2\overline{m}-1,2\overline{m^\prime}-1,\overline{\sigma}}(\vec{k},E) & \vec{k} \in \text{BZ}^\prime \\
\frac{s^2}{2}G_{2\overline{m},2\overline{m^\prime},\overline{\sigma}}(\vec{k}-\vec{Q},E) & \text{else}
\end{array}\right.
\end{align}
with $G$ the $4\times4$ matrix propagator from Eq.~\eqref{gdoublon}.

The $m,m^\prime$ layer index structure of $\Ac$ defines $\Ac_{11\sigma}=\Ac_{22\sigma}$ as the single-layer spectral function, whereas
\begin{align}
{\Ac}_\sigma(k_z) = \frac{1}{2}\!\left[{\Ac}_{11\sigma} + {\Ac}_{22\sigma} + {\rme}^{\rmi k_z}\!\left({\Ac}_{12\sigma} + {\Ac}_{21\sigma}\right)\right]
\end{align}
with $k_z \in \{0,\pi\}$ corresponds to the spectral function at fixed interlayer momentum $k_z$; the dependence on the spectator arguments $\vec{k},E$ has been suppressed for brevity. Note that all these spectra are independent of $\sigma$ as a consequence of time-reversal symmetry, so that we drop the index $\sigma$ when there is no risk of confusion.


\subsection{General aspects of the spectral function}

\begin{figure}
\centering
\includegraphics[width=0.375\textwidth]{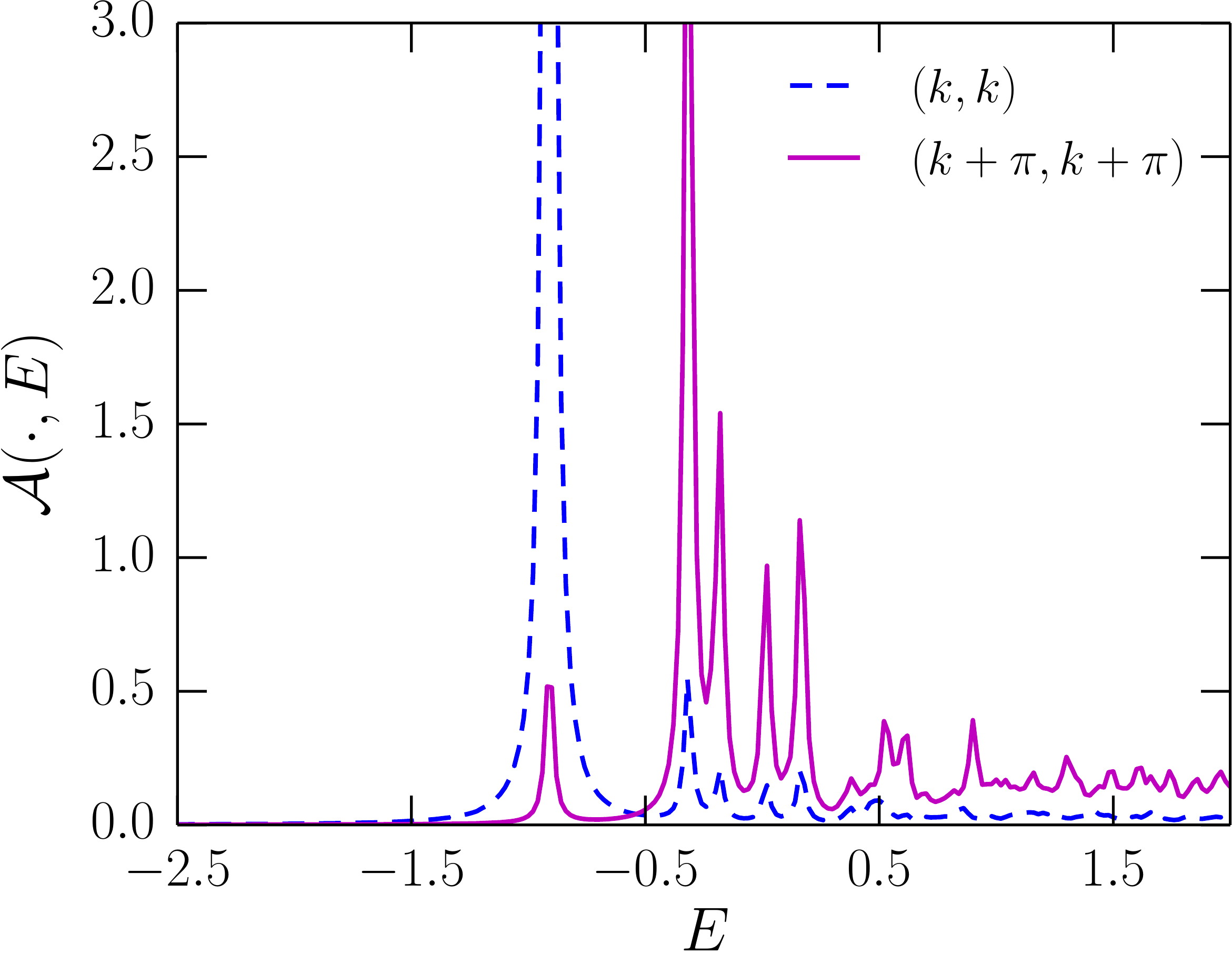}
\caption{
Spectral function $\mathcal{A}(\vec{k},E)$ for vanishing interlayer hopping, $t_\perp = 0$, at $\vec{k} = (k,k)$ for $k = 0.75\pi$ as well as its scattering partner $\vec{k} + \vec{Q}$, with hopping parameters $t = -0.5, t^\prime = 0, t^{\prime\prime} = -0.1$, and finite spin-orbit anisotropy $\theta = 12^\circ$. We choose $J_\perp = 3.06$ to set a triplon gap of $\Delta = 0.75$. All energies in units of $J$.
}
\label{fig:basicnonholt1}
\end{figure}

Before proceeding to actual \bslio{} phenomenology, we pause to reflect upon some general aspects of the spectral function. For our preliminary discussion, we consider a simplified setting without interlayer hopping ($t_\perp = 0$), with a set of hopping amplitudes such that $\theta = 0$ reduces to one of the $SU(2)$-symmetric single-layer scenarios considered in Refs.~\onlinecite{holt12,holt13}, and we have checked that our numerics reproduces the results therein (see also Appendix \ref{app:phtrafo} for the relation between hole and electron doping via particle-hole transformation).

In Fig.~\ref{fig:basicnonholt1}, we have nonzero spin-orbit anisotropy $\theta \neq 0$. The spectral functions are trivially diagonal and degenerate in layer index, and consist of coherent (Lorentzian) quasiparticle peaks and, in addition, an incoherent background comprising the many-particle continuum, which arises from the scattering of doublons via triplons. These constituents are generically found in spectral functions of fermions coupled to bosonic excitations. The crucial additional feature compared to the case without SOC-induced backfolding is the appearance of a shadow (the modifier ``shadow'' refers to the reduced intensity) peak at $\vec{k} + \vec{Q}$ (shown here for a nodal momentum point $\vec{k} = (k,k) \in \text{BZ}\setminus \text{BZ}^\prime$ in the outer half of the Brillouin zone, in the vicinity of where one would expect the experimentally relevant Fermi surface to be located) as a consequence of SOC: For $\theta \neq 0$, the translation symmetry is reduced and the unit cell is enlarged to $\sqrt{2}\times \sqrt{2}$, such that states at $\vec{k}$ and $\vec{k}+\vec{Q}$ couple. As a result, in fact all features in the spectrum at $\vec{k}$ display a ``shadow'' partner in the spectrum at $\vec{k}+\vec{Q}$, but with different relative intensities.

\begin{figure}
\centering
\includegraphics[width=0.375\textwidth]{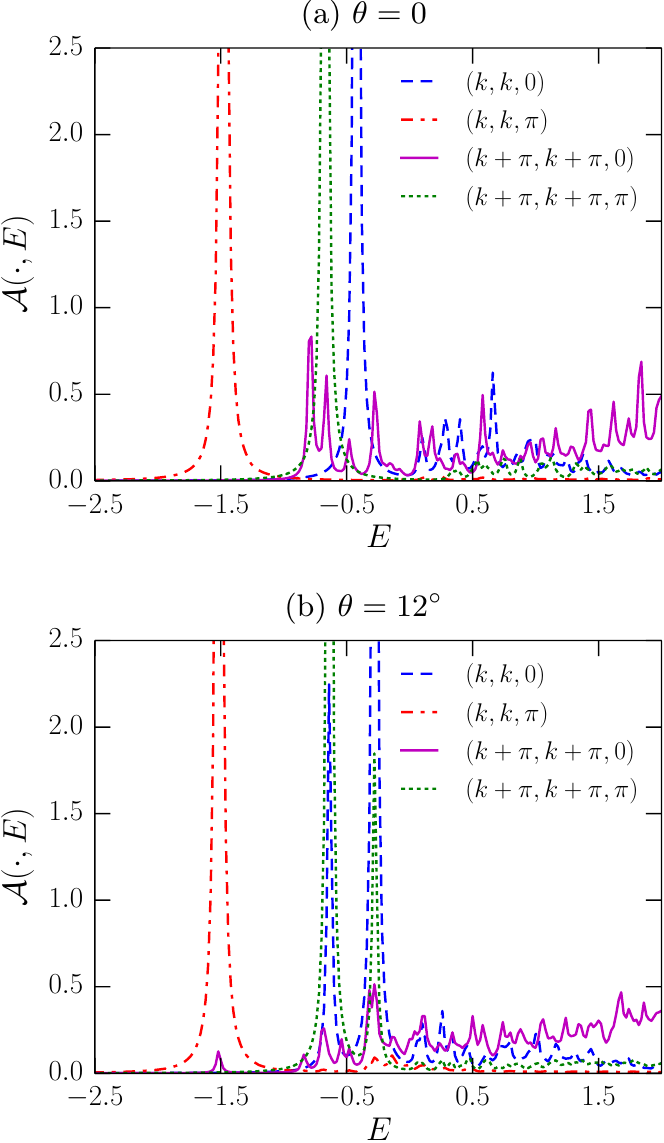}
\caption{Spectral function $\mathcal{A}(\vec{k},k_z,E)$ for $k = 0.75\pi$, with intralayer hopping parameters $t^{(n)}$ and triplon gap $\Delta$ as in Fig. \ref{fig:basicnonholt1}, but with finite interlayer coupling $t_\perp = 1.4|t|$. Panel (a) recapitulates the limit $\theta = 0$, equivalent to an SU(2)-symmetric model. Panel (b) shows, in addition, nonzero spin-orbit anisotropy $\theta = 12^\circ$, and thus contains the full complexity of the present problem.
}
\label{fig:basicnonholt2}
\end{figure}

When turning on interlayer coupling $t_\perp \neq 0$ (Fig.~\ref{fig:basicnonholt2}), it is convenient to carry out the discussion in the $k_z$-basis, since it is the eigenbasis of the effective theory ${\HH}_\text{eff}$ for $\theta = 0$. From the $SU(2)$-symmetric literature on spin ladders\cite{jureckabrenig01} and bilayers,\cite{vojtabecker98} it is known that one obtains two decoupled $k_z$-sectors, $k_z = 0$ and $k_z = \pi$, with the bilayer splitting $t_\perp$ between them, i.e. separating points in $(\vec{k},k_z)$-space with momentum difference $(0,0,\pi)$. This is reproduced for convenience (and as a consistency check) in Fig.~\ref{fig:basicnonholt2}(a). In fact, the energetic distance between the quasiparticle peaks at $(k,k,0)$ and $(k,k,\pi)$ allows one to read off a renormalized bilayer splitting. For $t_\perp>0$ the lowest-energy quasiparticle peak arises in the $k_z=\pi$ channel; this is the peak which will form the Fermi surface for finite doping.

In Fig.~\ref{fig:basicnonholt2}(b), we switch on spin-orbit anisotropy $\theta \neq 0$. In addition to bilayer splitting, one again finds, similar to the $t_\perp=0$ case of Fig.~\ref{fig:basicnonholt1}, the appearance of shadow peaks. Importantly, the shadows are accompanied by a momentum shift of $(\pi,\pi,\pi)$: The momentum transfer $q_z = \pi$ in the $z$ component arises due to the opposite staggering pattern in the different layers (recall that the phase factor was $\theta_{im\sigma} = \eta_m \theta_{i\sigma}$ in the hopping Hamiltonian ${\HH}_t$). We point out that this shadow phenomenology thus mimics that of the bilayer in the \emph{antiferromagnetic} phase (studied previously in Ref.~\onlinecite{vojtabecker98} for an SU(2)-symmetric model) despite the absence of static magnetic ordering; the role of the spatial structure of the magnetic order parameter is played by the pattern of the bond twists instead.


\subsection{Modeling \bslio}

Having discussed the general aspects of the spectral function, we now proceed to study the model with parameters relevant to \bslio{}, see Sec.~\ref{sec:para} for a discussion of the concrete parameter choice.

The ARPES experiments have been performed at a finite small level of doping, $x$, while our formalism is appropriate for the limit $x\to 0$. Neglecting the interaction between doublons, we can extrapolate to finite $x$ using a rigid-band approximation, i.e., by filling a Fermi sea of doublons. In practice, we consider constant-energy cuts through the spectral function $\Ac(\vec{k},E)$ to compare with the experimental photoemission intensity at the Fermi level. Here $E$ is chosen such that the momentum-space area $V_\text{Lut}$ enclosed by the quasiparticle peak matches the doping level $x$ according to $V_\text{Lut} = 3x/2$, with the factor $3/2$ coming from the stoichiometry of \bslio.

In order to compare spectral functions with ARPES data, we recall that only the momentum parallel to the surface of the sample, usually coinciding with the in-plane $2$-momentum $\vec{k}$, is conserved in an ARPES experiment. Furthermore, the escape depth of photoelectrons is limited. For simplicity, we therefore primarily focus on $\Ac_{11}(\vec{k},E)$ of the single layer, and comment on modifications due to a larger probing depth in Sec.~\ref{sec:shadowint} below.

\begin{figure*}
\includegraphics[width=0.75\textwidth]{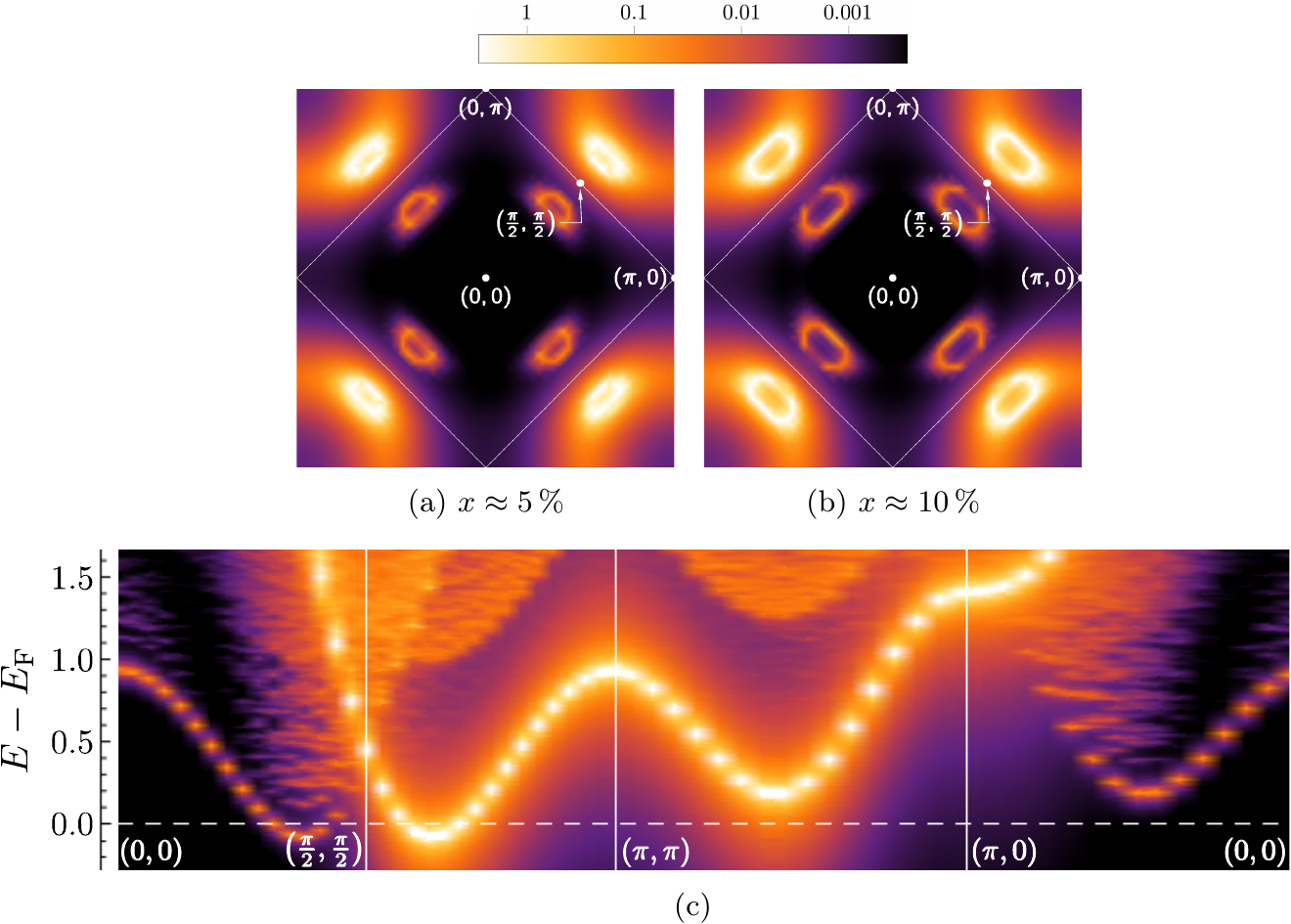}
\caption{Spectral function $\Ac_{11}$ for $t = -2, t^\prime = -0.7, t^{\prime\prime} = -0.6, t_\perp = 3.4, \theta = 12^\circ$ and $J_\perp = 2.96$ (in units such that $J = 1$), corresponding to a triplon gap $\Delta = 0.2$, with mean-field parameters $s^2 = 0.884$, $J_\perp/4 - \lambda = 3.59$. Panels (a) and (b) show constant-energy surfaces of the spectral function, with the corresponding doping level estimated from the Luttinger volume according to $V_\text{Lut} = 3x/2$, while (c) depicts the spectral function along a high-symmetry path of the Brillouin zone, with the Fermi energy $E_\text{F}$ corresponding to $x \approx10 \,\%$ (dashed line) as a guide to the eye.
}
\label{fig:mapA12+}
\end{figure*}

\subsubsection{Fermi surface at low doping}

Maps of the spectral function $\Ac_{11}$ for the main parameter set are displayed in Fig.~\ref{fig:mapA12+} (for concrete parameter values, see figure caption). In agreement with experiments such as those reported by de la Torre \textit{et al.},\cite{torre14} we find not only Fermi pockets centered around the nodal direction close to $(\pi/2,\pi/2)$,\cite{fn:tsign} but also, crucially, a---weaker in intensity---copy mirrored perpendicular to the nodal axis (termed ``shadow'') about the boundary of the reduced Brillouin zone. Likewise, one obtains a double pocket dispersion of the lowest quasiparticle band in energy--momentum space, cf. Fig. \ref{fig:mapA12+}(c). The appearance of the shadow band, as noted before, is a consequence of the interplay of SOC and octahedral rotation, which enlarges the unit cell and enables scattering between $\vec{k}$ and $\vec{k}+\vec{Q}$.
Compared to the ARPES data of Ref.~\onlinecite{torre14}, the Fermi surfaces arising from our calculations tend to be more extended in the nodal direction, and as a result less so in the direction perpendicular to it, especially at larger doping levels. Furthermore, the shadow intensity tends to be smaller than observed in experiments (see, however, Sec.~\ref{sec:shadowint} below). Nevertheless, the main qualitative experimental features are faithfully reproduced by the calculation. Although a less constrained fine-tuning of parameters (possibly along with the inclusion of more complicated hopping paths) might produce better agreement with experiment, we abstain from doing so here, since issues arising from physics beyond our approximations are just as likely to play a role when trying to obtain quantitative accuracy anyway.
\begin{figure}
\centering
\includegraphics[width=.45\textwidth]{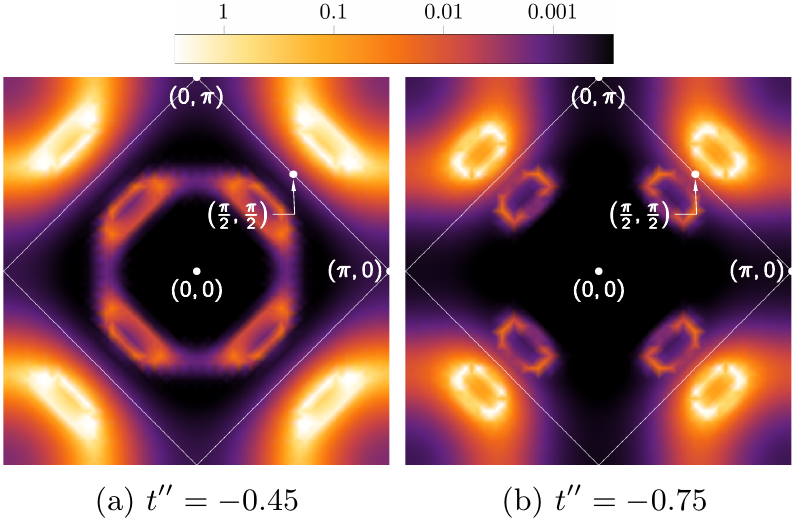}
\caption{Constant energy surfaces of the spectral function for (a) $t^{\prime\prime} = -0.45$ and (b) $t^{\prime\prime} = -0.75$, with other parameters as in Fig.~\ref{fig:mapA12+} and $E$ chosen to correspond to a doping level $x \approx10\,\%$.
}
\label{fig:mapA12-}
\end{figure}

\begin{figure}
\centering
\includegraphics[width=.45\textwidth]{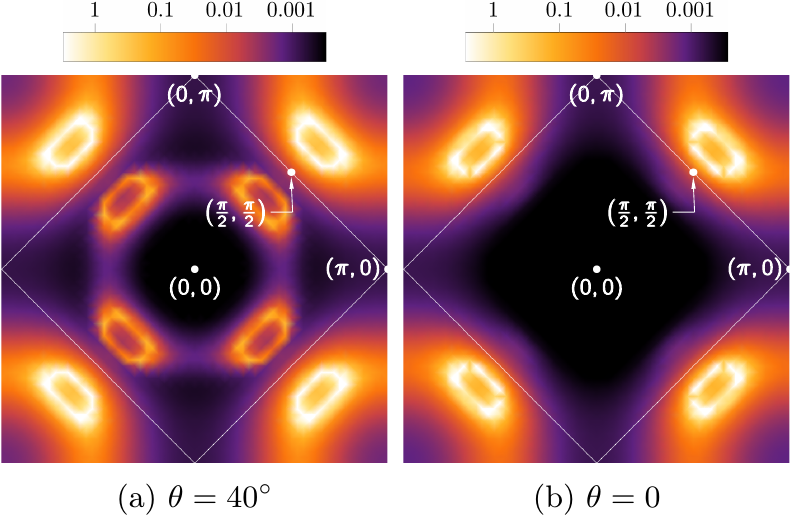}
\caption{Constant energy surfaces of the spectral function for (a) $\theta = 40^\circ$ and (b) $\theta = 0^\circ$, with other parameters as in Fig.~\ref{fig:mapA12+}, except for $J_\perp$ which was adjusted to fix the triplon gap at $\Delta = 0.2$. $E$ again corresponds to a doping level $x \approx10\,\%$.
}
\label{fig:mapA40}
\end{figure}

We now explore neighboring regions of parameter space, in order to discuss the significance of certain selected parameters. While the general fermiology is by and large robust, details are tunable to some degree by the value of longer-ranged hopping parameters $t^\prime/t$ and $t^{\prime\prime}/t$, see Fig.\ \ref{fig:mapA12-} for results. Roughly speaking, a smaller long-ranged hopping shifts the primary pockets away from the edge of the reduced Brillouin zone and towards $(\pm\pi,\pm\pi)$, while elongating it in the direction perpendicular to the nodal axis. The opposite trend occurs upon increasing $t^{\prime\prime}$.

So far, we have fixed the anisotropy parameter $\theta$ at a quite small value of $12^\circ$ in order to equal the actual rotation angle of the IrO$_6$ octahedra.\cite{subramanian94} Since deviations from this are possible depending on microscopic details of quantum chemistry,\cite{jin09,wangsenthil09} we now take the liberty of considering the opposite extreme, a rather large value of $\theta = 40^\circ$, in the range found by Moretti Sala \textit{et al.}\cite{moretti15} and Hogan \textit{et al.}\cite{hogan16} by fitting bond-operator theory at the harmonic approximation level to RIXS spectra. The primary effect is that a higher value of $\theta$ leads to a much higher intensity of the shadow pockets, cf. Fig.~\ref{fig:mapA40}(a). This behavior may in fact be anticipated at tree level, namely upon inspection of the quadratic part of the effective theory $H_{\psi\psi}$ \eqref{eq:hksigma}, since the terms with $(\tau^a)_{\kappa}$-structure ($a = 1,2$), i.e. responsible for mixing $\vec{k}$ and $\vec{k} + \vec{Q}$, scale $\propto \sin\theta$.

For comparison, we also show the case $\theta = 0$ case in Fig.~\ref{fig:mapA40}(b): This does not display any shadow Fermi surfaces, in agreement with previous studies on SU(2)-invariant models case,\cite{vojtabecker98,holt12,holt13} and confirms that the present shadows are generated by the combination of SOC and octahedral rotation.
It is worth recalling that shadow Fermi surfaces are observed experimentally in underdoped cuprates as well\cite{aebi94} where SOC is small; in this case the shadows are usually interpreted as precursors of antiferromagnetic order.\cite{kampfschrieffer90,monthoux95} This mechanism requires antiferromagnetic fluctuations at very small energies and is not relevant to our calculation.

A less pronounced (but still non-negligible) difference between Figs.~\ref{fig:mapA40}(a) and (b) lies in the location of the Fermi pockets, viz. the fact that the pockets are closer to the boundary of the reduced Brillouin zone for $\theta = 0$ than for $\theta \neq 0$. This is related to the lower symmetry of the magnetic sector in the presence of SOC and octahedral rotation: Concerning the location of renormalized quasiparticle peaks, the primary effect of the coupling to near-critical triplon modes in the presence of longer-ranged hopping is a shift of the dispersion minimum away from $(\pm\pi,\pm\pi)$ towards the center of the Brillouin zone.\cite{holt12,holt13} For $\theta = 0$, all triplon polarizations have a small gap close to quantum criticality. By contrast, for substantial $\theta$ only the $z$ mode gap is small, resulting in a smaller overall renormalization and pockets which are closer to $(\pm\pi,\pm\pi)$.

\subsubsection{Shadow intensity}
\label{sec:shadowint}

\begin{figure}
\centering
\includegraphics[width=0.45\textwidth]{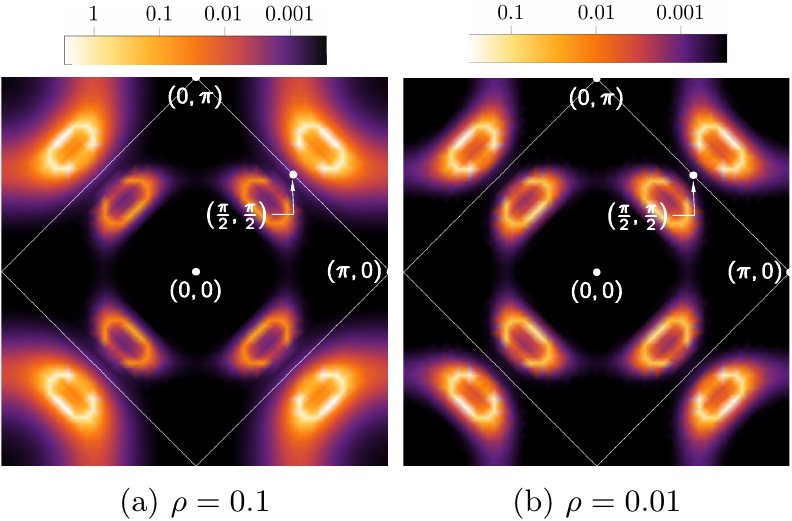}
\caption{Constant energy surfaces (Luttinger volume corresponding to $x \approx10\,\%$) of the spectral function $\Ac^{(\rho)}(\vec{k},E)$. All input parameters are the same as in Fig.\ \ref{fig:mapA12+}.}
\label{fig:Arho}
\end{figure}

We close this section by addressing the relative shadow intensity in the spectral function, recalling that the shadow and the main band occur in different $k_z$ channels. So far, we have considered $\Ac_{11}(\vec{k},E)$, corresponding to the signal from the upper layer (or equivalently an equal-weighted average of the $k_z = 0$ and $k_z = \pi$ channels), which led to a relatively weak shadow pocket signal.
A finite probing depth of ARPES tends to put a larger weight on $k_z = 0$,\cite{damascelliarpes} and hence we consider
\[
\Ac^{(\rho)}(\vec{k},E) = (1 - \rho)\,\Ac(\vec{k},0,E) + \rho\,\Ac(\vec{k},\pi,E)\;,
\]
where $\rho \in [0,1]$; the data in Fig.~\ref{fig:mapA12+} correspond to $\rho=1/2$. We focus on the case of small $\rho \ll 1$, which corresponds to a signal of predominantly $k_z = 0$ character. As shown in Fig. \ref{fig:Arho} for $\rho = 0.1$ and $0.01$ for the main parameter set (cf. Fig. \ref{fig:Arho}), the shadow intensity can easily be of the same order of magnitude as that of the main Fermi pocket. Since the probing depth depends on the photon energy in the ARPES experiment, we predict that the shadow intensity should display a significant dependence on photon energy.

\section{Summary and Outlook}

We have presented a theoretical description of the charge carrier dynamics in the paramagnetic phase of the bilayer iridate \bslio{}. We have employed a combination of bond operator mean-field theory and self-consistent Born approximation, applied to a suitable bilayer $t$--$J$ model, to calculate the single-particle spectral function at small carrier doping. Using a reasonable set of model parameters which placed the magnetic sector in the proximity to a bilayer quantum critical point, we were able to obtain a striking agreement with the Fermi surfaces measured in recent ARPES experiments, including the existence of doubled electron pockets. We have identified strong spin-orbit coupling in combination with the staggered rotation of IrO$_6$ octahedra---the main differences compared to previous work \cite{holt12,holt13}---as the origin of the shadow pockets, and we have predicted their relative intensity to depend strongly on ARPES photon energy.

The present description lends itself to a more detailed description of the physics at finite doping:
One may discuss the feedback of the charge carriers on the magnetic sector, i.e., the doping-induced renormalization of magnetic excitations, which has been studied experimentally in Ref.~\onlinecite{lu17}. This is left for future work.

More broadly, it would be interesting to elucidate the crossover from the doped Mott regime at small carrier concentration to a more conventional metallic regime at larger doping, cf. Fig.~\ref{fig:schempd}. Conceptually, this may indeed be a crossover as opposed to a transition, in contrast to the paramagnetic single-layer case relevant to cuprates: In the latter, a small-doping metallic state with Fermi pockets violates Luttinger's theorem and thus cannot be adiabatically connected to a conventional metal at large doping; instead, it has been proposed to realize a fractionalized Fermi liquid.\cite{flst1,moonss,ss16} In the bilayer case, the Fermi-pocket state does {\em not} violate Luttinger's theorem due to the doubled unit cell, and therefore no transition is required\cite{mv_af} to connect to a Fermi-liquid metal.


\acknowledgments

We acknowledge discussions with A. de la Torre, F. Mila, E. P\"arschke, and J. van den Brink. This research was supported by the DFG through SFB 1143 and GRK 1621.


\begin{appendix}

\section{Explicit expressions for effective doublon theory}
\label{app:expl}
In this appendix, we specify the terms appearing in the model of coupled doublons and triplons of Sec.~\ref{sec:scba}. The Hamiltonian matrix appearing in Eq.~\eqref{eq:efftheor-0} is given by
\begin{align}
\begin{split}
h_{\vec{k}\sigma} &= -2t s^2 \gamma_{\vec{k}}\left[\cos\theta\kern.05em(\tau^3)_{\kappa} \otimes (\mathbbm{1}_2)_m \right. \\
& \qquad \qquad \qquad \left. {} + \eta_\sigma\sin\theta\kern.05em(\tau^2)_{\kappa} \otimes (\tau^3)_m \right] \\
& \qquad {} - \left(2t^\prime \gamma^\prime_{\vec{k}} + 2t^{\prime\prime} \gamma^{\prime\prime}_{\vec{k}}\right)\!s^2 (\mathbbm{1}_2)_{\kappa} \otimes (\mathbbm{1}_2)_m \\
& \qquad {} + t_\perp (\mathbbm{1}_2)_{\kappa} \otimes (\tau^1)_m + \varepsilon_0  (\mathbbm{1}_2)_{\kappa} \otimes (\mathbbm{1}_2)_m\;,
\end{split}
\label{eq:hksigma}
\end{align}
where the longer-ranged form factors $\gamma^\prime,\gamma^{\prime\prime}$ are defined as
\begin{align*}
2\gamma^\prime_{\vec{k}} &= \cos(k_x + k_y) + \cos(k_x - k_y) \\
2\gamma^{\prime\prime}_{\vec{k}} &= \cos(2k_x) + \cos(2k_y)
\end{align*}
and $\varepsilon_0 = -E_0/N$ denotes the energy of a static doublon. Within the single-doublon sector, it only results in a constant shift of the energy and is hence disregarded in the calculations presented in the main text.
The interaction vertices $g_{\vec{k}\vec{q}\kern.05em{\kappa}\alpha\sigma\sigma^\prime}$ appearing in Eq.~\eqref{eq:efftheor-1} are $4\times4$ matrices (in spinor space) and have the form
\begin{align}
g_{\vec{k}\vec{q}\kern.05em {\kappa} \alpha \sigma\sigma^\prime} &= u_{\vec{q} {\kappa} \alpha} \kern.1em \tilde{g}_{\vec{k}\vec{q}\kern.05em{\kappa}\alpha\sigma\sigma^\prime} + v_{\vec{q} {\kappa} \alpha} \kern.1em \tilde{g}_{\vec{k}+\vec{q},\vec{q}\kern.05em{\kappa}\alpha\sigma\sigma^\prime}^{\kern.05em\dagger}\;,\nonumber
\end{align}
where $\tilde{g}_{\vec{k}\vec{q}\kern.05em{\kappa}\alpha\sigma\sigma^\prime}$ corresponds to the underlying (pre-Bogoliubov) interaction vertices of the form $t\psi^\dagger\psi$, which we parameterize in the following manner:
\begin{align}
\tilde{g}_{\vec{k}\vec{q}\kern.05em{\kappa}\alpha m m^\prime \sigma\sigma^\prime} &= \frac{s}{\sqrt{2}} \Bigl( \tilde{g}^{(t)}_{\vec{k}{\kappa} \alpha m \sigma} + \tilde{g}^{(J)}_{\vec{q} {\kappa} \alpha m \sigma} \Bigr)\tau^{3}_{mm^\prime}\tau_{\sigma^\prime\sigma}^\alpha\;, \nonumber
\end{align}
with the individual contributions from $\Ht,\HJ$ given by ($a = x,y$)
\begin{align*}
\tilde{g}^{(t)}_{\vec{p},+,zm\sigma} &= -2t\gamma_{\vec{p}}\kern-.1em\left(\cos \theta\kern.05em \tau^3 + \eta_m \eta_\sigma \sin \theta\kern.05em \tau^2 \right) \\
&\qquad - \left(2t^\prime \gamma^\prime_{\vec{p}} + 2t^{\prime\prime} \gamma^{\prime\prime}_{\vec{p}}\right) \mathbbm{1}_2, \\
\tilde{g}^{(t)}_{\vec{p},-,z m\sigma} &= 2 \rmi t\gamma_{\vec{p}}\kern-.1em\left(\cos \theta\kern.05em \tau^2 - \eta_m \eta_\sigma \sin \theta\kern.05em \tau^3 \right) \\
&\qquad - \left(2t^\prime \gamma^\prime_{\vec{p}} + 2t^{\prime\prime} \gamma^{\prime\prime}_{\vec{p}}\right)\kern-.1em\tau^1, \\
\tilde{g}^{(t)}_{\vec{p},+,am\sigma} &= 2t\gamma_{\vec{p}}\kern-.1em\left(\cos \theta\kern.05em \tau^3 - \eta_m \eta_\sigma \sin \theta\kern.05em \tau^2 \right) \\
&\qquad + \left(2t^\prime \gamma^\prime_{\vec{p}} + 2t^{\prime\prime} \gamma^{\prime\prime}_{\vec{p}}\right)\kern-.1em\mathbbm{1}_2, \\
\tilde{g}^{(t)}_{\vec{p},-,a m\sigma} &= -2\rmi t\gamma_{\vec{p}}\kern-.1em\left(\cos \theta\kern.05em \tau^2 - \eta_m \eta_\sigma \sin \theta\kern.05em \tau^1 \right) \\
&\qquad + \left(2t^\prime \gamma^\prime_{\vec{p}} + 2t^{\prime\prime} \gamma^{\prime\prime}_{\vec{p}}\right) \tau^1, \\
\tilde{g}^{(J)}_{\vec{p},+,zm\sigma} &= -J \gamma_{\vec{p}}\kern.05em \mathbbm{1}_2 \; , \; \tilde{g}^{(J)}_{\vec{p},-,zm\sigma} = J \gamma_{\vec{p}}\kern.05em \tau^1, \\
\tilde{g}^{(J)}_{\vec{p},+,am\sigma} &= J \gamma_{\vec{p}}\kern-.1em \left(\cos 2\theta\kern.05em \mathbbm{1}_2 \corr{{}+{}} \rmi \eta_m \eta_\sigma \sin 2\theta\kern.05em\tau^1\right), \\
\tilde{g}^{(J)}_{\vec{p},-,am\sigma} &= -J \gamma_{\vec{p}}\kern-.1em\left(\cos 2\theta\kern.05em \tau^1 \corr{{}-{}} \rmi \eta_m \eta_\sigma \sin 2\theta\kern.05em\mathbbm{1}_2\right),
\end{align*}
with all of them being $2\times2$ matrices in the $\kappa$ part of spinor space.


\section{Particle--hole transformation}
\label{app:phtrafo}
The present work deals with electron-doped Mott insulators, whereas the majority of previous works on carrier dynamics in magnets were motivated by cuprates and hence developed for hole doping. The two cases can be formally mapped onto each other using a particle--hole transformation $\hat{c},c \to \hat{a}^\dagger,a^\dagger$. Re-writing the Hamiltonian in terms of $a$ operators, $\Ht$ remains form-invariant upto a change of sign in the $t$ and $\theta$:
\begin{align}
&\Ht\bigl(\hat{c}_{im\sigma},\hat{c}^\dagger_{im\sigma};t^{(n)},\theta\bigr) = \Ht\bigl(\hat{a}_{im\sigma},\hat{a}^\dagger_{im\sigma};-t^{(n)},-\theta\bigr) \,.
\end{align}
On the present bipartite lattice, the sign change of $t,t_\perp$ can be absorbed in a shift of single-particle momenta by $(\pi,\pi,\pi)$. For the spin operators, we have
\[
S_{im\alpha} = \tfrac{1}{2}\tau^{\alpha}_{\sigma\sigma^\prime}\bigl(\delta_{\sigma\sigma^\prime} - a_{im\sigma^\prime}^\dagger\kern.05ema_{im\sigma}\bigr)\;.
\]
Thus, we have $\vec{S}_{im} = -\operatorname{diag}(1,-1,1)\,\vec{S}_{im}^{(a)}$, where
\[
S_{im\alpha}^{(a)} = \tfrac{1}{2}\kern.05em a_{im\sigma}^\dagger\kern.05em\tau^{\alpha}_{\sigma\sigma^\prime}\kern.05em a_{im\sigma^\prime}
\]
is the spin operator associated with the $a_{im\sigma}$. Under this transformation, the Heisenberg and pseudodipolar terms remain invariant, since they do not mix between different spin components, but the DM part picks up a minus sign, which may again be subsumed into the sign of $\theta$, i.e.
\begin{align}
{\HH}_J(\vec{S}_{im};J,\theta) = {\HH}_J\bigl(\vec{S}_{im}^{(a)};J,-\theta\bigr)\,.
\end{align}
Since the DM interaction does not enter the calculations in this paper (up to the order kept), we conclude that our results also apply for hole doping with flipped signs of $t^\prime,t^{\prime\prime}$ and after accounting for the momentum shift by $(\pi,\pi,\pi)$.

\end{appendix}



\begin{thebibliography}{99}

\bibitem{kim08} 
B. J. Kim, H. Jin, S. J. Moon, J.-Y. Kim, B.-G. Park, C. S. Leem, J. Yu, T. W. Noh, C. Kim, S.-J. Oh, J.-H. Park, V. Durairaj, G. Cao, and E. Rotenberg,
Phys. Rev. Lett. \textbf{101}, 076402 (2008).

\bibitem{moon08} 
S. J. Moon, H. Jin, K. W. Kim, W. S. Choi, Y. S. Lee, J. Yu, G. Cao, A. Sumi, H. Funakubo, C. Bernhard, and T. W. Noh,
Phys. Rev. Lett. \textbf{101}, 226402 (2008).

\bibitem{kim09} 
B. J. Kim, H. Ohsumi, T. Komesu, S. Sakai, T. Morita, H. Takagi, and T. Arima,
Science {\bf 323}, 1329 (2009).

\bibitem{jin09} H. Jin, H. Jeong, T. Ozaki, and J. Yu, %
Phys. Rev. B \textbf{80}, 075112 (2009).

\bibitem{witczakkrempa14} W. Witczak-Krempa, G. Chen, Y.~B. Kim, and L. Balents, %
Annu. Rev. Condens. Matter Phys. \textbf{5}, 57 (2014).

\bibitem{machida10} 
Y. Machida, S. Nakatsuji, S. Onoda, T. Tayama, and T. Sakakibara,
Nature \textbf{463}, 210 (2010).

\bibitem{li13} L. Li, P. P. Kong, T. F. Qi, C. Q. Jin, S. J. Yuan, L. E. DeLong, P. Schlottmann, and G. Cao, %
Phys. Rev. B \textbf{87}, 235127 (2013).

\bibitem{kim14} 
Y. K. Kim, O. Krupin, J. D. Denlinger, A. Bostwick, E. Rotenberg, Q. Zhao, J. F. Mitchell, J. W. Allen, and B. J. Kim,
Science \textbf{345}, 187 (2014).

\bibitem{cao16} 
Y. Cao, Q. Wang, J. A. Waugh, T. J. Reber, H. Li, X. Zhou, S. Parham, S.-R. Park, N. C. Plumb, E. Rotenberg, A. Bostwick, J. D. Denlinger, T. Qi, M. A. Hermele, G. Cao, and D. S. Dessau,
Nat. Commun. \textbf{7}, 11367 (2016).

\bibitem{leermp} P. A. Lee, N. Nagaosa, and X.-G. Wen,
Rev. Mod. Phys. {\bf 78}, 17 (2006).

\bibitem{moretti15} M. Moretti Sala, V. Schnells, S. Boseggia, L. Simonelli, A. Al-Zein, J. G. Vale, L. Paolasini, E. C. Hunter, R. S. Perry, D. Prabhakaran, A. T. Boothroyd, M. Krisch, G. Monaco, H. M. Ronnow D. F. McMorrow, and F. Mila, %
Phys. Rev. B \textbf{92}, 024405 (2015).

\bibitem{torre14} A. de la Torre, E. C. Hunter, A. Subedi, S. McKeown Walker, A. Tamai, T. Kim, M. Hoesch, R. Perry, A. Georges, and F. Baumberger,
Phys. Rev. Lett. \textbf{113}, 256402 (2014).

\bibitem{hogan16} T. Hogan, R. Dally, M. Upton, J. P. Clancy, K. Finkelstein, Y.-J. Kim, M. J. Graf, and S. D. Wilson,
Phys. Rev. B \textbf{94}, 100401 (2016).

\bibitem{lu17} 
X. Lu, D. E. McNally, M. Moretti Sala, J. Terzic, M. H. Upton, D. Casa, G. Ingold, G. Cao, and T. Schmitt,
Phys. Rev. Lett. \textbf{118}, 027202 (2017).

\bibitem{vojtabecker98} M. Vojta and K. W. Becker,
Phys. Rev. B \textbf{60}, 15201 (1998).

\bibitem{holt12} M. Holt, J. Oitmaa, W. Chen, and O. P. Sushkov,
Phys. Rev. Lett. \textbf{109}, 037001 (2012).

\bibitem{holt13} M. Holt, J. Oitmaa, W. Chen, and O. P. Sushkov,
Phys. Rev. B \textbf{87}, 075109 (2013).

\bibitem{mei12} J.-W. Mei,
arXiv:1210.1974.

\bibitem{fn:ba2iro4} Note that in a system without lattice anisotropies, such as Ba$_2$IrO$_4$, this has no further effect, and the hopping bilinear is trivial (i.e. proportional to the identity matrix).

\bibitem{wangsenthil09} F.~Wang and T.~Senthil,
Phys. Rev. Lett. \textbf{106}, 136402 (2011).

\bibitem{fn:units} We employ units such that $\hbar=1$ and the lattice constant $a=1$.

\bibitem{flst1} T. Senthil, S. Sachdev, and M. Vojta,
Phys. Rev. Lett. \textbf{90}, 216403 (2003).

\bibitem{moonss} 
E. G. Moon and S. Sachdev,
Phys. Rev. B \textbf{83}, 224508 (2011).

\bibitem{ss16} 
S. Sachdev, E. Berg, S. Chatterjee, and Y. Schattner,
Phys. Rev. B \textbf{94}, 115147 (2016).

\bibitem{plotnikova} E. M. P\"arschke, K. Wohlfeld, K. Foyevtsova, and J. van den Brink, %
Nat. Commun. \textbf{8}, 686 (2017).

\bibitem{fn:tsign} If the nearest-neighbor hopping had the opposite sign ($t>0$), then the primary Fermi pockets would be located inside the reduced Brillouin zone, in disagreement with experiment.\cite{torre14}

\bibitem{bil_split} L. Moreschini, S. Moser, A. Ebrahimi, B. Dalla Piazza, K. S. Kim, S. Boseggia, D. F. McMorrow, H. M. Rønnow, J. Chang, D. Prabhakaran, A. T. Boothroyd, E. Rotenberg, A. Bostwick, and M. Grioni
Phys. Rev. B \textbf{89}, 201114(R) (2014).

\bibitem{fn:tperpsign} The assumed sign of $t_\perp$ has no influence on the single-layer spectrum $\Ac_{11}$. Reversing the sign of $t_\perp$ exchanges the roles of $k_z=0$ and $\pi$, i.e., is relevant for the $k_z$-resolved spectrum.

\bibitem{chub95} A. V. Chubukov and D. K. Morr, Phys. Rev. B {\bf 52}, 3521 (1995).

\bibitem{kotovsushkov98} V. N. Kotov, O. Sushkov, Z. Weihong, and J. Oitmaa,
Phys. Rev. Lett. \textbf{80}, 5790 (1998).

\bibitem{matsushita99} Y. Matsushita, M. P. Gelfand, and C. Ishii, J. Phys. Soc. Jpn. \textbf{68}, 247 (1999).

\bibitem{sachdevbhatt90} S. Sachdev and R. N. Bhatt, Phys. Rev. B {\bf 41}, 9323 (1990).

\bibitem{fn:morettisalaHO} The harmonic approximation to bond-operator theory is obtained by inserting $t_{i0} = 1$ in \eqref{eq:bondrep}, i.e., assuming full singlet condensation as well as neglecting the hardcore constraint entirely. For the present spin-orbit anisotropic case (\bsio{}), this was done in Ref.~\onlinecite{moretti15}. Our improvement lies in the self-consistent treatment (at mean-field level) of the two aforementioned issues, viz. the singlet condensate parameter and the triplet hardcore constraint. The harmonic approximation can alternatively serve as the tree-level starting point of more refined diagrammatic approaches such as the Gell-Mann--Brueckner resummation method\cite{kotovsushkov98} or the $1/d$ expansion,\cite{joshivojta15a,joshivojta15b} which is beyond the scope of the present work.

\bibitem{sandvikscalapino94} A. W. Sandvik and D. J. Scalapino, Phys. Rev. Lett. {\bf 72}, 2777 (1994).

\bibitem{jureckabrenig01} C. Jurecka and W. Brenig,
Phys. Rev. B \textbf{63}, 094409 (2001).

\bibitem{fn:triplonbip} To go from the $\HJtwo$ in \eqref{eq:HJtwohelpers}--\eqref{eq:Bogovars} to the bipartite version, replace $\gamma(\vec{q}) \to \kappa \gamma(\vec{q})$ in the definition of $B_\alpha$ and all subsequent formul\ae{}.

\bibitem{schmittrink88} S. Schmitt-Rink, C. M. Varma, and A. E. Ruckenstein, %
Phys. Rev. Lett. \textbf{60}, 2793 (1988).

\bibitem{kaneleeread89} C. L. Kane, P. A. Lee, and N. Read, %
Phys. Rev. B \textbf{39}, 6880 (1989).

\bibitem{eder98} R. Eder, %
Phys. Rev. B \textbf{57}, 12832 (1998).

\bibitem{altlandsimons} A.~Altland, and B.~Simons, \textit{Condensed Matter Field Theory}, 2nd ed., Cambridge University Press, Cambridge (2010).

\bibitem{fn:structureG} More explicitly, the $\sigma$-blockdiagonal structure of the Green's function (at tree level a consequence of the blockdiagonal form of $h_\sigma$) is preserved in the SCBA approximation because the interaction vertices are either blockdiagonal ($\alpha = z$) or only comprise off-diagonal blocks ($\alpha = x,y$).

\bibitem{subramanian94} M. A. Subramanian, M. K. Crawford, and R. L. Harlow, %
Mater. Res. Bull. \textbf{29}, 645 (1994).

\bibitem{aebi94} P. Aebi, J. Osterwalder, P. Schwaller, L. Schlapbach, M. Shimoda, T. Mochiku, and K. Kadowaki, %
Phys. Rev. Lett. \textbf{72}, 17 (1994)

\bibitem{kampfschrieffer90} A. P. Kampf, and J. R. Schrieffer, %
Phys. Rev. B \textbf{42}, 7967 (1990).

\bibitem{monthoux95} P. Monthoux,
Phys. Rev. B \textbf{55}, 11111 (1995).

\bibitem{damascelliarpes} A.~Damascelli, %
Phys. Scr. \textbf{T109}, 61 (2004).

\bibitem{mv_af} M. Vojta,
Phys. Rev. B {\bf 78}, 125109 (2008).

\bibitem{joshivojta15a} D. G. Joshi, K. Coester, K. P. Schmidt, and M. Vojta,
Phys. Rev. B \textbf{91}, 094404 (2015).

\bibitem{joshivojta15b} D. G. Joshi and M. Vojta,
Phys. Rev. B \textbf{91}, 094405 (2015).


\end{thebibliography}
\end{document}